\documentclass[njp,10pt]{iopart}
\usepackage{graphicx}
\expandafter\let\csname equation*\endcsname\relax
\expandafter\let\csname endequation*\endcsname\relax
\usepackage{amsmath}
\usepackage{amsfonts}
\usepackage{amssymb}
\usepackage{soul}
\usepackage{dsfont}
\usepackage{hyperref}
\usepackage{amstext}
\usepackage{braket}
\usepackage[caption=false]{subfig}
\usepackage{braket}

\date{\today}

\begin{document}

\title[Enhanced continuous generation of non-Gaussianity through optomechanical mod...]{Enhanced continuous generation of non-Gaussianity through optomechanical modulation}

\author{Sofia Qvarfort$^{1}$, Alessio Serafini$^{1}$, Andr\'e Xuereb$^{2}$, Dennis R\"atzel$^{3}$, David Edward Bruschi$^{4,5}$}
\address{$^{1}$Department of Physics and Astronomy, University College London, Gower Street, WC1E 6BT London, United Kingdom}
\address{$^{2}$Department of Physics, University of Malta, Msida MSD 2080, Malta}
\address{$^{3}$Institut f\"{u}r Physik, Humboldt-Universit\"{a}t zu Berlin, 12489 Berlin, Germany}
\address{$^{4}$Faculty of Physics, University of Vienna, Boltzmanngasse 5, 1090 Vienna, Austria}
\address{$^{5}$ Institute for Quantum Optics and Quantum Information - IQOQI Vienna, Austrian Academy of Sciences, Boltzmanngasse 3, 1090 Vienna, Austria}
\ead{sofiaqvarfort@gmail.com, david.edward.bruschi@gmail.com}

\begin{abstract}
We study the non-Gaussian character of quantum optomechanical systems evolving under the fully nonlinear optomechanical Hamiltonian. By using a measure of non-Gaussianity based on the relative entropy of an initially Gaussian state, we quantify the amount of non-Gaussianity induced by both a constant and time-dependent cubic light--matter coupling and study its general and asymptotic behaviour. We find analytical approximate expressions for the measure of non-Gaussianity and show that initial thermal phonon occupation of the mechanical element does not significantly impact the non-Gaussianity. 
More importantly, we also show that it is possible to continuously increase the amount of non-Gaussianity of the state by driving the light--matter coupling at the frequency of mechanical resonance, suggesting a viable mechanism for increasing the non-Gaussianity of optomechanical systems even in the presence of noise. 
\end{abstract}
\maketitle

\section{Introduction}\label{intro}
Understanding nonlinear, interacting physical systems is paramount across many areas in physics. Specifically, ``nonlinear'' (or ``anharmonic'') dynamical systems include all those whose Hamiltonian cannot be expressed as a second-order polynomial in the quadrature operators. Crucially, these systems allow us to generate non-Gaussian states, which cannot be done given only quadratic couplings. One family of system where this is possible are optomechancial systems, where light interacts with a mechanical element through a cubic interaction term. 

In recent years, the intrinsic value of nonlinear systems, as opposed to the aforementioned limitations that linear systems face, has been made clearer and more rigorous. It has been shown that non-linearities in the form of non-Gaussian states constitute an important resource for quantum teleportation protocols~\cite{dell2010teleportation}, universal quantum computation~\cite{lloyd1999quantum, menicucci2006universal}, quantum error correction~\cite{niset2009no}, and entanglement distillation~\cite{eisert2002conditions, fiuravsek2002gaussian, giedke2002characterization}. This view of non-Gaussianity as a resource for information-processing tasks has inspired recent work on developing a resource theory based on non-Gaussianity~\cite{zhuang2018resource, takagi2018convex, albarelli2018resource}. In addition, it has been found that non-Gaussianity provides a certain degree of robustness in the presence of noise~\cite{sabapathy2011robustness, nha2010linear}. 

In the context of quantum information and computation, there has been a drive towards the realisation of anharmonic Hamiltonians as well as more general methods and control schemes capable of generating and stabilising non-Gaussian states~\cite{silberhorn2001generation, lvovsky2001quantum,  heersink2003polarization, ourjoumtsev2006quantum, ourjoumtsev2007generation, parigi2007probing}. 
On the one hand, this is motivated by the fact that, in order to obtain effective qubits from the truncation of infinite dimensional systems, one needs unevenly spaced energy levels, such that only the transition between the two selected energy levels may be targeted and driven. In turn, this requires a sufficiently anharmonic Hamiltonian. On the other hand, it has always been clear that protocols entirely restricted to Gaussian preparations, manipulations and read-outs, through quadratic Hamiltonians and general-dyne detection, are classically simulatable, as their Wigner functions 
may be mimicked by classical probability distributions~\cite{mari2012positive}.\footnote{We should note here that, since uncertainties in quantum Gaussian systems are fundamentally bounded by the Heisenberg uncertainty relation, which in principle does not hold for classical systems, Gaussian operations are in fact sufficient to run some protocols requiring genuine quantum features, such as continuous variable quantum key distribution.}

In optomechanical systems~\cite{aspelmeyer2014cavity}, where electromagnetic radiation is coherently coupled to the motion of a mechanical oscillator, the light--matter interaction induced by radiation pressure is inherently non-linear~\cite{mancini1997ponderomotive, bose1997preparation, ludwig2008optomechanical}. The nonlinear features of optomechanical systems have been frequently explored in the context of metrology, such as force sensing~\cite{caves1980measurement} and gravimetry~\cite{qvarfort2018gravimetry, armata2017quantum}. In particular, the nonlinear coupling enables the creation of optical cat states in the form of superpositions of coherent states~\cite{mancini1997ponderomotive, bose1997preparation}. These cat states can also be transferred to the mechanics~\cite{palomaki2013coherent}, which opens up the possibility of using massive superpositions for testing fundamental phenomena such as collapse theories~\cite{goldwater2016testing} and, potentially, signatures of gravitational effects on quantum systems at low energies~\cite{bose2017spin,marletto2017gravitationally}. In addition to cat states, other non-Gaussian states such as compass states~\cite{zurek2001sub, toscano2006sub} and hypercube states~\cite{howard2018hypercube},  and also all been found to have excellent sensing capabilities.
This combination of sensing with nonlinear state and fundamental applications makes it imperative to explore the nonlinear properties of the mechanical systems.

\begin{figure}[ht!]
\centering
  \includegraphics[width=.7\linewidth, trim = 14mm 0mm 10mm 12mm]{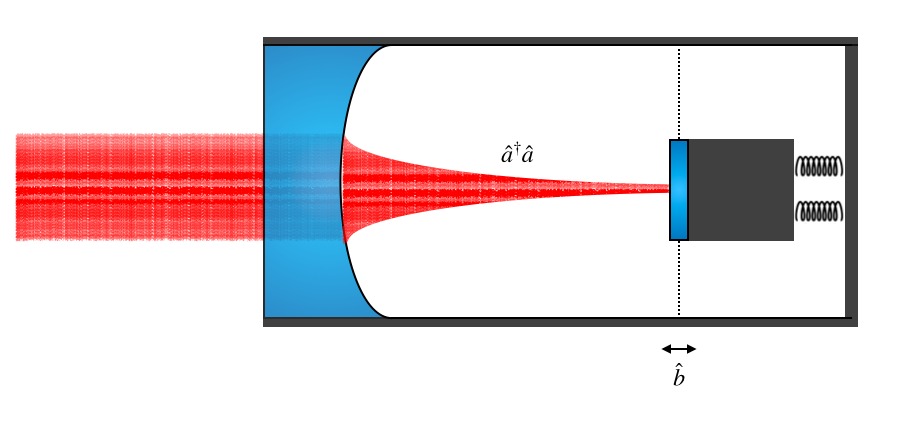}%
\caption{An optomechanical system consisting of a moving end mirror. The operators $\hat{a}$ and $\hat{a}^\dag$ denote the creation and annihilation operators for the light field, and $\hat{b}$ and $\hat{b}^\dag$ denote the motional degree of freedom of the mirror. Other examples of optomechanical systems include levitated nanobeads and cold--atom ensembles.  }
\label{fig:optomechanics}
\end{figure}

A number of different optomechanical systems have been experimentally implemented, including Fabry--P\'{e}rot cavities with a moving-end mirror~\cite{favero2009optomechanics}, as shown in Figure \ref{fig:optomechanics}, levitated nano-diamonds~\cite{barker2010cavity, yin2013large}, membrane-in-the-middle configurations~\cite{jayich2008dispersive} and optomechanical crystals~\cite{eichenfield2009picogram, safavi2014two}. While several experiments have demonstrated genuine nonlinear behaviour (see for example~\cite{sankey2010strong, doolin2014nonlinear,brawley2016nonlinear,  leijssen2017nonlinear}), most experimental settings can however be fully modelled with linear dynamics~\cite{kippenberg2007cavity, leijssen2017nonlinear}, and it is generally difficult to access the fully nonlinear regime. As a result, significant effort has been devoted to the question of how the non-linearity can be further enhanced. Most approaches focus on the few-excitation regime, where increasing the inherent light--matter coupling allows for detection of the non-linearity. This enhancement can be achieved, for example, by using a large-amplitude, strongly detuned mechanical parametric drive~\cite{lemonde2016enhanced}, or by modulating the spring constant~\cite{yin2017nonlinear}. Similar work has shown that the inclusion of a mechanical quartic anharmonic term can be nearly optimally detected with homodyne and heterodyne detection schemes, which are standard measurements implemented in the laboratory~\cite{latmiral2016probing}. 

A natural question that arises considering the approaches above  is: \textit{Are there additional methods by which the amount of non-Gaussianity in an optomechanical system can be further increased}? One such proposal was put forward in~\cite{liao2014modulated} where it was suggested that the nonlinearity in electromechanical systems could be enhanced by several orders of magnitude by modulating the light--matter coupling. This is achieved by driving the system close to mechanical resonance and takes a simple form in the rotating-wave approximation. Here, we seek to fully quantify the non-Gaussianity of the exact, nonlinear optomechanical state for both ideal and open systems. More precisely, given an initial Gaussian state evolving under the standard optomechanical Hamiltonian, we quantify how non-Gaussian the state becomes as a function of time and the parameters of the Hamiltonian in question. To do so, we make use of recently developed analytical techniques to study the time-evolution of time-dependent systems~\cite{Bruschi:Xuereb:2018}, and employ a specific measure of non-Gaussianity based on the relative entropy of the state~\cite{genoni2008quantifying}. Our results include the fact that the non-Gaussianity of an optomechanical system initially in a coherent state scales with the inherent light--matter coupling, as expected. We also find that the non-Gaussianity scales logarithmically with the coherent state parameter of the optical system, and we illustrate how this behaviour differs for small and large coherent states. 
Most importantly, however, we find that the non-Gaussianity of the state can be continuously increased by modulating the light--matter coupling strength at the mechanical resonant frequency. Such a continuous increase might prove especially useful given open system dynamics, where it would allow for the creation of an effectively non-Gaussian steady state. Moreover, we find that the system does not have to be cooled to the ground state in order to access significant amounts of the non-Gaussianity. Finally, we discuss several methods by which this modulation can be realised for optomechanical systems in the laboratory. 

This paper is organised as follows. In Section~\ref{tools} we present the Hamiltonian of interest and solve the resulting dynamics. In Section~\ref{sec:measure:non:gaussianity} we introduce the measure of non-Gaussianity from~\cite{genoni2008quantifying}, and in Section~\ref{section:results} we derive some generic results based on the measure which apply in different regimes. We then proceed to examine the behaviour of the non-Gaussianity in optomechanical systems for two cases: a constant light--matter coupling in Section~\ref{sec:constant:coupling}; and a time-dependent coupling in Section~\ref{sec:time:dependent:coupling}, where we also show that driving the coupling results in continuously generated non-Gaussianity. Both preceding sections also include an analysis of the open system dynamics. Finally, in Section~\ref{sec:discussion} we discuss our results and propose various methods by which the modulation of the optomechanical coupling can be achieved. Section~\ref{sec:conclusions} concludes this work, and many of the derivations used in this work can be found in ~\ref{appendix:one},~\ref{appendix:one:time:evolution},~\ref{section:symplectic:eigenvalues}, and~\ref{app:time:dependent:coefficients}

\section{Dynamics}\label{tools}
We start our work by presenting the necessary mathematical tools needed to solve the dynamics. We refer the reader to~\ref{appendix:one} for a more extensive introduction on the techniques presented below. 

\subsection{Hamiltonian}\label{optomech}
We begin by considering two bosonic modes corresponding to an electromagnetic mode and a mechanical oscillator. Without loss of generality, we shall henceforth refer to the electromagnetic mode as the optical mode. The operators $\hat a$ and $\hat b$ of the cavity and mechanical modes respectively, obey the canonical commutation relations $[\hat a,\hat a^\dag]=[\hat b,\hat b^\dag]=1$, while all other commutators vanish. The radiation pressure induces a nonlinear interaction between the light and mechanics, and the whole systems is modelled by the following Hamiltonian:
\begin{align}\label{main:time:independent:Hamiltonian:to:decouple}
	\hat {H} &= \hbar\,\omega_\mathrm{c}\,\hat a^\dagger \hat a + \hbar\,\omega_\mathrm{m} \,\hat b^\dagger \hat b - \hbar\,g(t)\,\hat a^\dagger\hat a \bigl(\hat b+ \hat b^\dagger\bigr),
\end{align}
where $\omega_\mathrm{c}$ and $\omega_\mathrm{m}$ are the frequencies of the cavity mode and the mechanical mode respectively, and $g(t)$ drives the, potentially time-dependent, nonlinear light--matter coupling. The light--matter coupling strength $g(t)$ takes on different functional forms for different optomechanical systems, and we also note that this Hamiltonian governs the evolution of many similar systems, including electro--optical systems~\cite{tsang2010cavity}.

To simplify our notation and expressions, we rescale the laboratory time $t$ by the frequency $\omega_\mathrm{m}$, therefore introducing the dimensionless time $\tau:=\omega_\mathrm{m}\,t$, the dimensionless frequency $\Omega:=\omega_\mathrm{c}/\omega_\mathrm{m}$, and the dimensionless coupling $\tilde{g}(\tau):=g(t\omega_\mathrm{m})/\omega_\mathrm{m}$. This choice will prove convenient throughout the rest of this work, and dimensions can be restored when necessary by multiplying by $\omega_{\mathrm{m}}$. This renormalisation effectively is equivalent to the use of time $\tau$ and the Hamiltonian
 \begin{align}\label{main:time:independent:Hamiltonian:to:decouple:dimensionless}
	\hat {H}/(\hbar \omega_{\mathrm{m}}) &= \Omega \,  \hat a^\dagger \hat a + \hat b^\dagger \hat b -\tilde{g}(\tau) \hat a^\dagger\hat a \bigl(\hat b+ \hat b^\dagger\bigr).
\end{align}
To determine the action of this Hamiltonian on initial states, we now proceed to solve the dynamics induced by~\eqref{main:time:independent:Hamiltonian:to:decouple:dimensionless}. 

\subsection{Time evolution of the system} \label{subsec:time:evolution}
We now need an expression for the time evolution operator $\hat{U}(\tau)$ for a system evolving with~\eqref{main:time:independent:Hamiltonian:to:decouple:dimensionless}. The unitary time-evolution operator reads 
\begin{equation}\label{time:evolution:operator}
\hat{U}(\tau):=\overset{\leftarrow}{\mathcal{T}}\,\exp\biggl[-i \, \int_0^{\tau} d\tau'\,\hat{H}(\tau')\biggr],
\end{equation}
where $\overset{\leftarrow}{\mathcal{T}}$ is the time-ordering operator~\cite{Bruschi:Lee:2013}. This expression simplifies dramatically when the Hamiltonian $\hat{H}$ is independent of time, in which case one simply has $\hat{U}(\tau)=\exp\bigl(-i\,\hat{H}\,\tau \bigr)$. As we will here consider time-dependent light--matter couplings $\tilde{g}(\tau)$, we instead seek to solve the full dynamics of the time-dependent Hamiltonian. 

To do so, we make the ansatz that the time-evolution operator can be written as a product of a number of operator $\hat{U}_j$. This is possible if there exists a finite set of generators $\hat{G}_j$ that form a closed Lie algebra under commutation~\cite{Bruschi:Lee:2013}. We thus write the evolution operator~\eqref{time:evolution:operator} as 
\begin{equation}
\hat{U}(\tau) = \prod_j \hat{U}_j (\tau) = \prod_j e^{- i F_j (\tau)\, \hat{G}_j}, 
\end{equation}
where $F_j(\tau)$ are generally time-dependent coefficients determining the influence of the generator $\hat{G}_j$ on the quantum state. Our task is to find these coefficients, which has been done in ~\cite{Bruschi:Xuereb:2018} for an analogous setting. We note that compared with~\cite{Bruschi:Xuereb:2018}, the operators $\hat{a}$ and $\hat{b}$ have been swapped and that in our case the coupling $\tilde{g}(\tau)$ is preceded by a minus-sign.  For clarity, we therefore present the full derivation for the case considered here in~\ref{appendix:one}. 

With these techniques, we find that the time-evolution operator $\hat{U}(\tau)$ can be cast into the convenient form 
\begin{align}\label{explicit:time:evolution:operator}
\hat U(t)=e^{-i\,\hat{N}_b\,\tau}\,e^{-i\,F_{\hat{N}^2_a}\,\hat{N}^2_a}\,e^{-i\,F_{\hat{N}_a\,\hat{B}_+}\,\hat{N}_a\,\hat{B}_+} e^{-i\,F_{\hat{N}_a\,\hat{B}_-}\,\hat{N}_a\,\hat{B}_-}, 
\end{align} 
where the operators, given by, 
\begin{align}\label{basis:operator:Lie:algebra}
	\hat{N}_a &:= \hat a^\dagger \hat a &
	\hat{N}_b &:= \hat b^\dagger \hat b 
	 & \hat{N}^2_a &:= (\hat a^\dagger \hat a)^2\nonumber\\
	\hat{N}_a\,\hat{B}_+ &:= \hat{N}_a\,(\hat b^{\dagger}+\hat b) &
	\hat{N}_a\,\hat{B}_- &:= \hat{N}_a\,i\,(\hat b^{\dagger}-\hat b), &
	 & 
\end{align}
form a closed Lie algebra under commutation,  and where the coefficients that determine the evolution in~\eqref{explicit:time:evolution:operator} are given by 
\begin{align}\label{sub:algebra:decoupling:solution:text}
F_{\hat{N}^2_a}&= 2  \,\int_0^\tau\,\mathrm{d}\tau'\,\tilde{g}(\tau')\,\sin(\tau')\int_0^{\tau'}\mathrm{d}\tau''\,\tilde{g}(\tau'')\,\cos(\tau''),\nonumber\\
F_{\hat{N}_a\,\hat{B}_+}&=- \int_0^\tau\,\mathrm{d}\tau'\,\tilde{g}(\tau')\,\cos(\tau'),\ \text{and}\nonumber\\
F_{\hat{N}_a\,\hat{B}_-}&= -\int_0^\tau\,\mathrm{d}\tau'\,\tilde{g}(\tau')\,\sin(\tau').
\end{align}
Note that in~\eqref{explicit:time:evolution:operator}, we have transformed into a frame rotating with $\Omega_{\mathrm{c}} \, \hat{N}_a$ in order to neglect the phase-term $e^{- i \Omega \tau}$.  Given an explicit form of $\tilde{g}(\tau)$, it is then possible to write down a full solution for $\hat{U}(\tau)$. The decoupling techniques necessary to obtain this compact solution have a long tradition in quantum optics~\cite{Puri:2001} and were generalised and refined recently~\cite{Bruschi:Lee:2013}.
Finally, before we proceed, we also define the following two parameters:
\begin{align} \label{eq:combined:coefficients}
\theta_a &= 2 \, \left( F_{\hat{N}_a^2} + F_{\hat{N}_a \, \hat{B}_+} F_{\hat{N}_a \hat{B}_-} \right) \nonumber \\
F &= F_{\hat{N}_a \, \hat{B}_-} + i F_{\hat{N}_a \, \hat{B}_+}. 
\end{align}
These quantities will be useful when discussing features of the non-Gaussianity. 

\subsection{Recovering the standard dynamics}
Let us show here that this method reproduces the standard evolution operator for the optomechanical Hamiltonian. For this specific case, the light--matter interaction is held constant with $\tilde{g}(\tau) =\tilde{g}_0$. 
The functions~\eqref{sub:algebra:decoupling:solution:text} simplify to 
\begin{align} \label{eq:coefficients:time:independent:text}
F_{\hat N_a^2 } &=- \tilde{g}_0^2  \, \bigl[1-\textrm{sinc}(2\tau)\bigr]\,\tau,  \nonumber \\
F_{\hat N_a \, \hat B_+} &= - \tilde{g}_0  \,  \sin{( \tau )},  \nonumber \\
F_{\hat N_a \, \hat B_-} &= \tilde{g}_0 \, \bigl[ \cos{(\tau)}  -1\bigr].
\end{align}
These coefficients allow us to write the time evolution operator $\hat U(\tau)$  as
\begin{equation} \label{plain:optomechanics:evolution:operator:text}
\hat U(\tau)= e^{- i \,\hat{N}_b\,\tau} \, e^{ i\,\tilde{g}_0^2\,[1 - \textrm{sinc}(2\tau)]\,\tau\,\hat{N}_a^2}\,e^{i\,\tilde{g}_0\,\sin{(\tau)}\, \hat{N}_a\,\hat{B}_+}\,e^{i\,\tilde{g}_0\,(1 - \cos{(\tau)})\,\hat{N}_a\,\hat{B}_-} .
\end{equation}
This expression matches that found in the literature (see e.g. equation 3 in ~\cite{bose1997preparation}, which can be obtained with some rearrangement of the terms in~\eqref{plain:optomechanics:evolution:operator:text}). 

\subsection{Initial states of the system} \label{sec:initial:state}
In this work, we will examine the non-Gaussianity of the evolved state given two initial states: a coherent state and a thermal coherent state. 

\begin{itemize}
	\item[i)] \textbf{Coherent states.} We start by considering the case when both the optical and the mechanical modes are in a coherent state, which we denote $\ket{\mu_\mathrm{c}}$ and $\ket{\mu_\mathrm{m}}$ respectively. These states satisfy the relations $\hat{a}\ket{\mu_\mathrm{c}}=\mu_\mathrm{c}\ket{\mu_\mathrm{c}}$ and $\hat{b}\ket{\mu_\mathrm{m}}=\mu_\mathrm{m}\ket{\mu_\mathrm{m}}$. For the optical field, this is a readily available resource, since coherent states model laser light quite well. The mechanical element in optomechanical systems is most often found in a thermal state or, assuming perfect preliminary cooling, in its ground state, with $\mu_\mathrm{m} = 0$.
The initial state $\ket{\Psi(0)}$ of the compound system will therefore be
\begin{equation}\label{initial:state:two}
\ket{\Psi(0)} = \ket{\mu_{\mathrm{c}}} \otimes \ket{\mu_{\mathrm{m}}}. 
\end{equation} 
	\item[i)] \textbf{Thermal coherent states.} The assumption that the mechanics is in the ground state is not always justified, and therefore we shall also consider the non-Gaussianity of cases where the mechanics is in a thermal coherent state~\cite{vanner2011pulsed}. Such state is obtained simply by integrating over the coherent state parameter with an appropriate kernel~\cite{barnett2002methods}. We define the thermal state $\hat \rho_{\mathrm{th}}$ as 
\begin{equation}
\hat \rho_{\mathrm{th}} = \frac{1}{\bar{n} \pi} \int \mathrm{d}^2 \beta \, e^{- |\beta|^2/\bar{n}} \ket{\beta}\bra{\beta}, 
\end{equation} 
where $\bar{n}$ is the average thermal phonon occupation of the state, $\ket{\beta}$ is a coherent state,  and the integration occurs over the full complex space. Assuming that the optical mode in the coherent state $\ket{\mu_{\mathrm{c}}}$, the full initial state is therefore given by 
\begin{equation} \label{initial:coherent:thermal:state}
\hat \rho (0) = \ket{\mu_{\mathrm{c}}}\bra{\mu_{\mathrm{c}}} \otimes \hat \rho_{\mathrm{th}}
\end{equation}
\end{itemize}

By starting in an initial Gaussian state, we ensure that any non-Gaussianity revealed by our work is due to the nonlinear coupling in Eq.~\eqref{main:time:independent:Hamiltonian:to:decouple:dimensionless}. Indeed, the only way an initially Gaussian state may at any time be non-Gaussian is for the corresponding Hamiltonian to induce some nonlinear evolution~\cite{Adesso:Ragy:2014}. We do however note that while the measure of non-Gaussianity that we shall make use of has a clear and operational notion of the measure for pure states, it is harder to make statements about the non-Gaussianity of states that are mixed, such as the coherent thermal state $\hat{\rho}_{\mathrm{th}}$. See Section~\ref{sec:discussion} for a discussion of the properties of the relative entropy measure. 

\subsection{Open system dynamics} \label{subsec:noisy}
All realistic systems experience decoherence. In optomecahnical systems, this manifests as photons leaking from the cavity or as damping of the mechanical motion. Given a sufficiently weakly coupled environment, we can model the open dynamics of the system with the help of the Lindblad equation~\cite{gardiner2004quantum}. We note, however, that there is increasing evidence that the standard master equation treatment breaks down, especially in the strong coupling regime \cite{naseem2018thermodynamic}. Here, we shall only consider weak coupling in the presence of noise, and so the Lindblad equation is given by 
\begin{equation}
\dot{\hat{\rho}} = - i \left[ \hat H , \hat \rho \right] + \hat L_{\mathrm{c}} \, \hat \rho \, \hat L_{\mathrm{c}}^\dag  +\hat L_{\mathrm{m}} \, \hat \rho \, \hat L_{\mathrm{m}}^\dag - \frac{1}{2} \{ \hat L_{\mathrm{c}}^\dag \hat{L}_{\mathrm{c}}, \hat \rho \}- \frac{1}{2} \{ \hat L_{\mathrm{m}}^\dag \hat{L}_{\mathrm{m}}, \hat \rho \}, 
\end{equation}
where $\hat L _{\mathrm{c}}$ and $\hat L_{\mathrm{m}}$ are the Lindblad operators for the optics and mechanics, respectively. To model photon and phonon decay, we assume that $\hat L_{\mathrm{c}} = \sqrt{\kappa_{\mathrm{c}}} \, \hat a$ and $\hat L_{\mathrm{m}} = \sqrt{\kappa_{\mathrm{m}}} \, \hat b$, where $\kappa_{\rm{c}}$ is the optical decoherence rate and $\kappa_{\rm{m}}$ is the phonon decoherence rate. 

While analytic solutions for this particular choice of $\hat L_{\mathrm{m}}$ were obtained in~\cite{bose1997preparation}, photon decay can currently only be modelled numerically. We therefore make use of the Python library \textit{QuTiP} to simulate the noisy state evolution and its effect on the non-Gaussianity of the resulting state. We shall examine the non-Gaussianity for open systems in Sections~\ref{sec:constant:coupling} and~\ref{sec:time:dependent:coupling}, but first, we  define the measure of non-Gaussianity. 

\section{Measures of deviation from Gaussianity} \label{sec:measure:non:gaussianity}
Given a Hamiltonian $\hat{H}$, and an initial \textit{Gaussian} state $\hat \rho(0)$, we ask the following question: \textit{can we quantify how much the state $\hat \rho(\tau)$ deviates from a Gaussian state at time $\tau$}? This question stems from the following observation. The dynamics of our system is non-linear. Therefore, we expect an initial Gaussian state, characterised by a Gaussian Wigner function, to become a non-Gaussian state at later times. In fact, the only way for a Gaussian state to preserve its Gaussian character would be to evolve through a \textit{linear} transformation, which is induced by a Hamiltonian with at most quadratic terms in the quadrature operators~\cite{serafini2017quantum}.

To answer our question we first need to find a suitable measure of deviation from Gaussianity. In this work we choose to employ a measure for pure states, which we denote $\delta$, that is based on the comparison between the entropy of the final state and that of a suitably chosen reference Gaussian state~\cite{genoni2008quantifying}. A similar measure has been used to compute features of mixed systems~\cite{park2018faithful}. 

\subsection{Measures of deviation from Gaussianity: definition} \label{sec:measure:non:gaussianity:definition}
Let us detail here the construction of the non-Gaussianity quantifier $\delta(\tau)$ for our nonlinear dynamics. First, our initial state $\hat{\rho}(0)$ evolves into the state $\hat{\rho}(\tau)$ at time $\tau$. With our full solutions for the dynamics in Section~\ref{subsec:time:evolution}, we can find analytic expressions for the first and second moments of $\hat{\rho}(\tau)$. Then, we construct a state $\hat{\rho}_\textrm{G}(\tau)$,  which is the Gaussian state defined by the first and second moments that coincide with those of $\hat{\rho}(\tau)$. Now, we recall that a Gaussian state is \textit{fully} defined by its first and second moments. Therefore, if two Gaussian states $\hat{\rho}_1$ and $\hat{\rho}_2$ have equal first and second moments  they are the same state~\cite{Adesso:Ragy:2014,serafini2017quantum}. In general, our state $\hat{\rho}(\tau)$ at time $\tau$ will not be Gaussian and therefore cannot be specified fully by its first and second moments. This implies that we can introduce a measure $\delta(\tau)$ that quantifies how $\hat{\rho}(\tau)$ deviates from $\hat{\rho}_\textrm{G}(\tau)$:
\begin{equation}\label{measure:of:non:gaussianity}
\delta(\tau):=S(\hat{\rho}_\textrm{G}(\tau))-S(\hat{\rho}(\tau)),
\end{equation}
where $S(\hat{\rho})$ is the von Neumann entropy of a state $\hat{\rho}$, defined by $S(\hat{\rho}):=-\Tr(\hat{\rho}\,\ln\hat{\rho})$. This measure has been shown to capture the intrinsic non-Gaussianity of the system, and it vanishes if and only if $\hat{\rho}(\tau)$ is a Gaussian state~\cite{genoni2008quantifying}. In other words, if at all times $\delta(\tau) = 0$ this implies that the state is Gaussian and  the dynamics is fully linear. 

We now note that the time evolution is unitary. This means that $S(\hat{\rho}(\tau))=S(\hat{\rho}(0))$. If we start from a pure state, then $S(\hat{\rho}(0))=0$ and $\delta(\tau)=S(\hat{\rho}_\textrm{G}(\tau))$. We discuss the case where the initial state is mixed in Section~\ref{sec:discussion}.  

\subsection{Measures of deviation from Gaussianity: computation using the covariance matrix formalism} \label{sec:measure:non:gaussianity:cmformalism}
Since $\hat{\rho}_\textrm{G}(\tau)$ is a Gaussian state, its entropy can be exactly computed using the covariance matrix formalism~\cite{Adesso:Ragy:2014,serafini2017quantum}. The covariance matrix consists of the second moments of a quantum state, and can be used to fully characterise a Gaussian state (along with its first moments).  This is convenient, as the construction of $\hat{\rho}_{\mathrm{G}}$ involves finding the first of second moments of $\hat{\rho}$ anyway. While we could compute the entropy for $\hat{\rho}_{\mathrm{G}}$  by finding a diagonal basis in the Hilbert space, there exists a straight-forward method within the covariance matrix formalism. 

To compute the entropy, we introduce the $4\times4$ covariance matrix $\boldsymbol{\sigma}(\tau)$ of the state $\hat{\rho}_\textrm{G}(\tau)$. This matrix contains the second moments of the state $\hat{\rho}(\tau)$ which in our specific choice of basis $\hat{\mathbb{X}} = (\hat{a} , \hat{b} ,\hat{a}^\dag, \hat{b}^\dag)^{\mathrm{T}}$ is defined through its elements $\sigma_{nm}(\tau)  := \langle \{\hat{X}_n,\hat{X}_m^\dag\}\rangle-2\,\langle\hat{X}_n\rangle\langle\hat{X}_m^\dag\rangle$, 
where $\{\bullet,\bullet\}$ is the anti-commutator, we have defined the expectation value of an operator $\langle\bullet\rangle:=\Tr\{\bullet\,\rho_{\textrm{G}}\}$, and where for the sake of simpler notation, we have chosen not to write out the time-dependence explicitly. To compute the entropy $S(\hat{\rho}_{\textrm{G}}(\tau))$ we require the symplectic eigenvalues $\{\pm\nu_+(\tau),\pm\nu_-(\tau)\}$ of $\boldsymbol{\sigma}(\tau)$, where the property $\nu_\pm(\tau)\geq1$ holds for all states. The symplectic eigenvalues can be computed by finding the eigenvalues of the object $i\,\boldsymbol{\Omega}\,\boldsymbol{\sigma}(\tau)$, where $\boldsymbol{\Omega}$ is the symplectic form given by $\boldsymbol{\Omega}=\textrm{diag}(-i,-i,i,i)$ in this basis. The von Neumann entropy $S(\boldsymbol{\sigma})$ is then given in this formalism by $S(\boldsymbol{\sigma}) = s_V(\nu_+)+s_V(\nu_-)$, where the binary entropy $s_V(x)$ is defined by $s_V(x) := \frac{x + 1}{2} \ln \left( \frac{x+1}{2} \right) - \frac{x - 1}{2} \ln \left( \frac{x - 1}{2} \right)$. In summary, the state $\hat{\rho}(\tau)$ is non-Gaussian at time $\tau$ if and only if $\delta(\tau)>0$.

\subsection{Measures of deviation from Gaussianity: General behaviour} \label{subsec:general:behaviour}
We shall now infer some general characteristics of the measure of non-Gaussianity $\delta(\tau)$. The following analysis only holds when the system is pure, which in our case means that we assume that both the optics and mechanics start off as coherent states. As mentioned above, the symplectic eigenvalues $\nu_\pm$ satisfy $\nu_\pm\geq1$~\cite{serafini2017quantum}. Therefore, we can conveniently write $\nu_\pm=1+\delta\nu_\pm$, where $\delta \nu_\pm\geq0$ captures any deviation from purity. If the state is pure at time $\tau=0$ then it follows that $\nu_\pm(0)=1$~\cite{serafini2017quantum}. If the evolution is linear, then it is also the case that $\nu_\pm(\tau)=1$ for all $\tau$. For closed dynamics, the symplectic eigenvalues may only change if the evolution is non-linear. In this case, we would define $\nu_\pm=\nu_{0\pm}+\delta\nu_\pm$ with $\nu_{0\pm}>1$. Then, we would have that $\nu_\pm(0)=\nu_{0\pm}$ and, again, linear evolution would imply that $\nu_\pm(\tau)=\nu_{0\pm}$. The preceding statements imply that $\delta\nu_\pm$ are functions of the nonlinear contributions alone. Thus, when the non-linearity tends to vanish, then $\delta\nu_\pm\rightarrow0$. Among the possible asymptotic regimes we have that $\delta\nu_\pm\rightarrow+\infty$ or that it becomes constant.

These observations are important. To understand their implications we use the expression $\nu_\pm=1+\delta\nu_\pm$ to compute the general deviation from Gaussianity as $\delta(\tau)=s_V(1+\delta\nu_+)+s_V(1+\delta\nu_-)$. Using this form, we see that in the nearly linear (Gaussian) regime with only small contributions from the nonlinear dynamics, we will have $\delta\nu_\pm\ll1$ and therefore
\begin{align}\label{general:measure:of:non:gaussianity:small}
\delta(\tau)\approx&-\frac{\delta\nu_+}{2}\ln\frac{\delta\nu_+}{2}-\frac{\delta\nu_-}{2}\ln\frac{\delta\nu_-}{2}.
\end{align}
On the contrary, in the highly nonlinear (non-Gaussian) regime we have $\delta\nu_\pm\gg1$ and therefore $\delta(\tau)\approx\ln\frac{\delta\nu_+}{2}+\ln\frac{\delta\nu_-}{2}$.
If the symplectic eigenvalues depend on a large parameter $x\gg1$, then one will in general find that they have the asymptotic form $\nu_\pm\sim x^{N_\pm}\sum^{N_\pm}_{n=0}\nu_\pm^{(n)}\,x^{-n}$ for some appropriate real coefficients $\nu_\pm^{(n)}$, where $N_{\pm}$ constitutes the upper limits of the sum~\cite{ahmadi2014quantum}. 

A careful asymptotic expansion of the measure of nonlinearity in this regime gives
 \begin{equation}\label{large:parameter:prediction:expansion}
 \delta(\tau)\sim (N_++N_-)\,\ln x.
\end{equation}
These general results allow us to anticipate that these general behaviours will be confirmed by the explicit analytical and numerical computations below. A detailed computation can be found in~\ref{section:symplectic:eigenvalues}.

\section{General results } \label{section:results}
We now proceed to evolve the initial state $\hat \rho(0)$ with the evolution operator $\hat{U}(\tau)$ in~\eqref{explicit:time:evolution:operator}. 
To compute the amount of non-Gaussianity of this state $\delta(\tau)$, we must first find the elements $\sigma_{nm}$ of the covariance matrix $\boldsymbol{\sigma}$, which we construct form the first and second moments of 
$\hat{\rho}(\tau)$. We do so in full generality, meaning that the light--matter coupling $\tilde{g}(\tau)$ can take the form of any time-dependent function. 

We have computed the second moments and the covariance matrix $\boldsymbol{\sigma}$ in~\ref{appendix:one:time:evolution}. The second moments for the mechanical being in an initial coherent state and in an initial coherent thermal state can be found in~\eqref{app:expectation:values:coherent} and~\eqref{app:expectation:values:coherent:thermal} respectively. These can then be used to compute the covariance matrix elements $\sigma_{nm}$ where $n,m$ take values $0,1,2,3$. We have explicitly computed the elements of $\mathbf{\sigma}$ for a mechanical coherent state in~\eqref{full:elements:covaraince:matrix}. 

Our challenge now is to compute the symplectic eigenvalues $\nu_\pm$, given the expressions~\eqref{full:elements:covaraince:matrix}. The process of computing the eigenvalues can be simplified by using the expression $2\,\nu_\pm^2=\Delta\pm\sqrt{\Delta^2-4\,\textrm{det}(\boldsymbol{\sigma})}$, which is based on the existence of symplectic invariants~\cite{serafini2017quantum}. The definition of $\Delta$ is given in~\ref{appendix:one:two}. The full analytic expression for the symplectic eigenvalues $\nu_\pm$ of $\boldsymbol{\sigma}$ in~\eqref{full:elements:covaraince:matrix} is too long and cumbersome to be printed here. It also does not yield any immediate insight into the behaviour of $\delta(\tau)$. Instead, we will proceed to derive two analytic expressions for the two different regimes we identified in Section~\ref{subsec:general:behaviour}; small and large coherent state parameters respectively.

\subsection{Asymptotic behaviour for a small optical coherent state parameter}
We begin by looking at the case where $|\mu_{\mathrm{c}}|^2\ll1$ and where the mechanics is initially in a coherent state. Here, one can take the covariance matrix elements~\eqref{full:elements:covaraince:matrix} and, after some algebra, show that the perturbative expansion of the symplectic eigenvalues gives
\begin{align} 
\nu_+\sim&1+\left(1-|F|^2\,e^{-|F|^2}\right)\,|\mu_{\mathrm{c}}|^2\nonumber\\
\nu_-\sim&1+\left(1-e^{-|F|^2}\right)\,|F|^2\,|\mu_{\mathrm{c}}|^2.
\end{align}
This implies that the behaviour of $\delta(\tau)$ for small $|\mu_{\mathrm{c}}|$ goes as 
\begin{equation} \label{eq:delta:small}
\delta(\tau) \sim -\left(1+\left(1-2\,e^{-|F|^2}\right)\,|F|^2\right)\,|\mu_{\mathrm{c}}|^2\,\ln|\mu_{\mathrm{c}}|,
\end{equation}
in perfect agreement with~\eqref{general:measure:of:non:gaussianity:small}. This approximation suggests that $\delta(\tau)$ scales with $\sim |F|^2 |\mu_{\mathrm{c}}|^2 \ln |\mu_{\mathrm{c}}|$ to leading order. 

These expressions do not hold if the mechanical element is initially mixed. However, we will find in the next section that initial phonon occupation only marginally affects the non-Gaussianity.

\subsection{Asymptotic behaviour for a large optical coherent state parameter} \label{sec:asymptotic:large:coherent:state}

We now investigate the case where $|\mu_\mathrm{c}|\gg1$. Our goal is to derive an analytic expression for the non-Gaussianity that can be used to analyse the overall features of $\delta(\tau)$. Before making any quantitative evaluation, we recall that the measure will have the form (\ref{large:parameter:prediction:expansion}), where now $x\equiv|\mu_\mathrm{c}|^2$. Let us proceed to demonstrate this result analytically for this specific case. 

For large $\mu_{\mathrm{c}}$ and for the mechanics in the ground-state $\mu_{\mathrm{m}} = 0$, it is clear that whenever $\theta_a \neq 2\pi n $ for integer $n$, the matrix elements $\sigma_{31}$, $\sigma_{21}$ and $\sigma_{41}$ in~\eqref{full:elements:covaraince:matrix} vanish, due to the exponentials containing the  factor $|\mu_{\mathrm{c}}|^2$. Therefore, far (enough) from the times where $\theta_a = 2\pi n$  we are left with the following covariance matrix elements
\begin{align}\label{covariance:matrix:elements:large:input:photons:far:from:critical:times}
\sigma_{11} &\sim \sigma_{33} = 1 + 2|\mu_{\mathrm{c}}|^2 \left( 1 - e^{-4 \, |\mu_{\mathrm{c}}|^2 \sin^2{\theta_a/2}} \, e^{- |F|^2}  \right) \nonumber \\
\sigma_{22} &\sim \sigma_{44} = 2 \, |\mu_{\mathrm{c}}|^2 |F|^2 + 1\nonumber \\
\sigma_{42} &\sim \sigma_{24}^* =2 \, |\mu_{\mathrm{c}}|^2  \, e^{- 2i \tau} F^{*2},
\end{align}
and all other elements  are zero. We have kept the full expression for $\sigma_{11}$ because it reproduces some key elements of $\delta$ which we shall discuss later. Therefore, we do not expect the thermal occupation of the mechanics to significantly affect the non-Gaussianity that can be accessed in this system. Note also that we need to keep the next leading order in each element of~\eqref{covariance:matrix:elements:large:input:photons:far:from:critical:times}, which is a constant in the case of $\sigma_{11}$ and $\sigma_{22}$. Naively neglecting of this element would give an incorrect result when computing the entropy, as the neglected factor becomes significant in the logarithm~\cite{ahmadi2014quantum}. If the thermal element is in a coherent thermal state, however, the expectation values of $\braket{a}$ changes slightly and $\sigma_{11}$ in~\eqref{covariance:matrix:elements:large:input:photons:far:from:critical:times} will look different.  However, if we approximate $\sigma_{11}$ as $\sigma_{11} \approx 1  + 2 \, |\mu_{\mathrm{c}}|^2$, which follows from that $\braket{a} \sim 0 $ for very large $\mu_{\mathrm{c}}$, the non-zero covariance matrix elements of the coherent thermal mechanical state are the same as in~\eqref{covariance:matrix:elements:large:input:photons:far:from:critical:times}. We therefore conclude that an initially thermal coherent mechanical state will also exhibit most of the non-Gaussianity we will examine for coherent mechanical states. 

With this simplified matrix, we are able to find a simple and analytic expression for the symplectic eigenvalues, which reads
\begin{align} \label{eq:approx:symplectic:eigenvalues}
\nu_+ &\sim 1 + 2|\mu_{\mathrm{c}}|^2 \left( 1 - e^{-4 \, |\mu_{\mathrm{c}}|^2 \sin^2{\theta_a/2}} \, e^{- |F|^2}  \right) \nonumber \\
\nu_- &\sim\sqrt{4|\mu_{\mathrm{c}}|^2 \, |F|^2 +1} .
\end{align}
We note that both eigenvalues grow with $|\mu_{\mathrm{c}}|$, as expected from our analysis in Section~\ref{subsec:general:behaviour}. The amount of non-Gaussianity for large $\mu_{\mathrm{c}}$ is now given by the following expression
\begin{equation} \label{eq:delta:large}
\delta(\tau) \sim  s_V \left( 1 + 2|\mu_{\mathrm{c}}|^2 \left( 1 - e^{-4 \, |\mu_{\mathrm{c}}|^2 \sin^2{\theta_a/2}} \, e^{- |F|^2}  \right)\right) + s_V(\sqrt{4|\mu_{\mathrm{c}}|^2 \, |F|^2 +1}),
\end{equation}
which scales asymptotically as $\delta(\tau)\sim\tilde{\delta}(\tau) :=  4\ln\,|\mu_{\mathrm{c}}|$, in perfect agreement with~\eqref{large:parameter:prediction:expansion}.

Note that~\eqref{eq:delta:large} is also valid for a time-dependent light--matter coupling $\tilde{g}(\tau)$. In all cases, the nonlinearity grows as $\ln{|\mu_{\mathrm{c}}|}$ to leading order. In Sections~\ref{subsec:large:coherent:state} and~\ref{subsec:large:coherent:state:resonance} we will compare the asymptotic measure $\tilde{\delta}(\tau)$ with the full measure $\delta$ for different cases.

\section{Applications: Constant coupling} \label{sec:constant:coupling}
Let us now move on to a quantitative analysis of the evolving non-Gaussianity in different contexts. 
We begin by considering the case where the nonlinear light--matter interaction is constant: $\tilde{g}(\tau)=\tilde{g}_0$. To a large extent, this is the case for most experimental systems. The coefficients  which determine the time-evolution are those found in~\eqref{eq:coefficients:time:independent:text}, and we note that the function $F$, defined in~\eqref{eq:combined:coefficients}, which appears in the covariance matrix elements $\sigma_{nm}$ is now given by $F = \tilde{g}_0 \, ( 1 - e^{- i \tau})$. 

\begin{figure}[t!]
\centering
\subfloat[ \label{fig:measure:vs:time:various:mu}]{%
  \includegraphics[width=.42\linewidth, trim = 14mm 0mm 10mm 12mm]{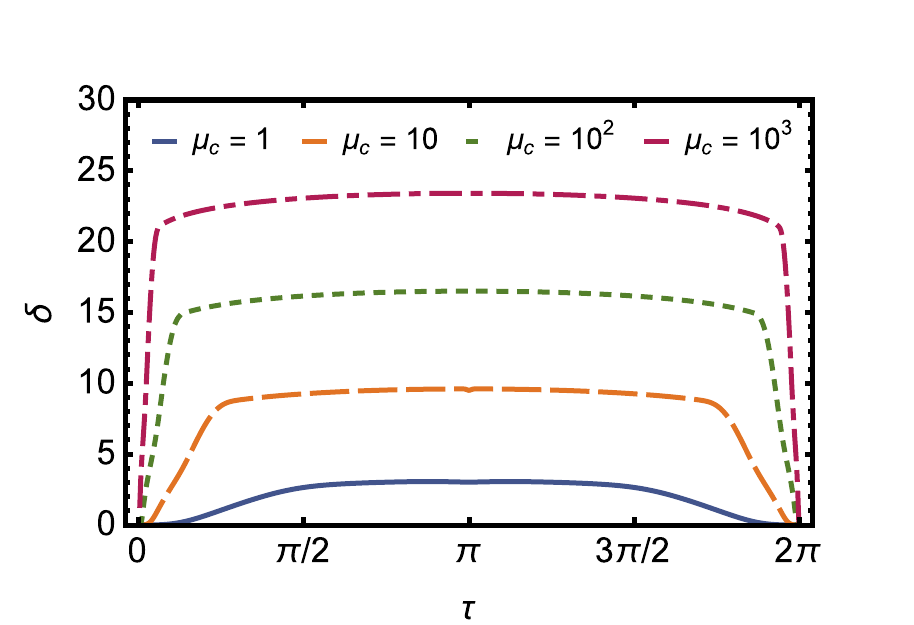}%
}\hfill
\subfloat[ \label{fig:measure:vs:time:various:mu:around:pi}]{%
  \includegraphics[width=.42\linewidth, trim = 14mm 0mm 10mm 12mm]{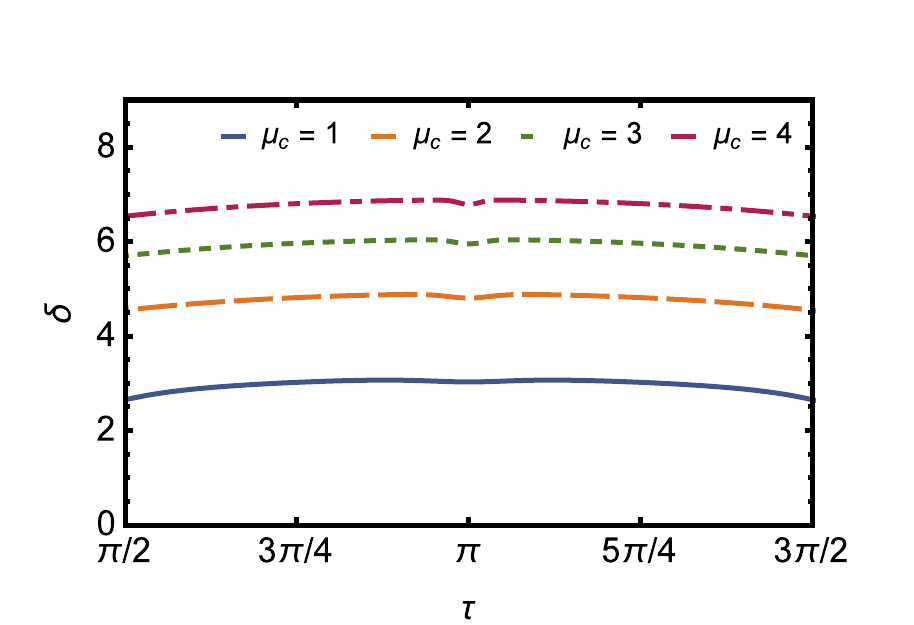}%
}\hfill
\subfloat[
\label{fig:measure:for:small:tau}]{%
  \includegraphics[width=.42\linewidth, trim = 14mm 0mm 10mm 12mm]{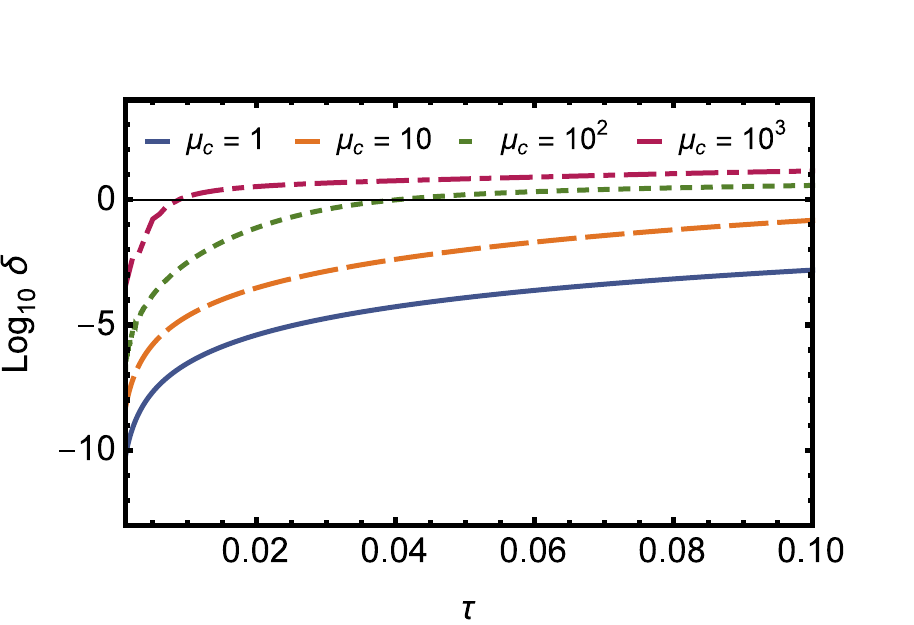}%
}\hfill
\caption{The measure of non-Gaussianity  $\delta(\tau)$ vs. time $\tau$ for systems with constant nonlinear coupling $\tilde{g}_0$. \textbf{(a)} A plot of $\delta(\tau)$ as a function of time $\tau$ for different coherent state parameters $\mu_{\mathrm{c}}$. The rescaled coupling is $\tilde{g}_0 = 1$ and the mechanics is in the ground state with  $\mu_{\mathrm{m}} = 0$. \textbf{(b)} A plot of $\delta$ vs. time $\tau$ near $\tau = \pi$ for varying  $\mu_{\mathrm{c}}$. The measure displays a local minimum centered around $\tau = \pi$ that becomes sharper with larger $\mu_{\mathrm{c}}$. Here $\tilde{g}_0 = 1$ and $\mu_{\mathrm{m}} = 0$. \textbf{(c)} A plot of $\log_{10} \delta(\tau)$ at very small times $\tau$ for different coherent state parameters $\mu_{\mathrm{c}}$, $\tilde{g}_0 = 1$ and $\mu_{\mathrm{m}} = 0$. The measure increases exponentially at first before it plateaus towards a constant value, which is the overall behaviour we observe in~(a). }
\label{fig:constant:measure:vs:time}
\end{figure}

\begin{figure}[t!]
\subfloat[
\label{fig:measure:vs:C1}]{%
  \includegraphics[width=.42\linewidth, trim = 14mm 0mm 10mm 12mm]{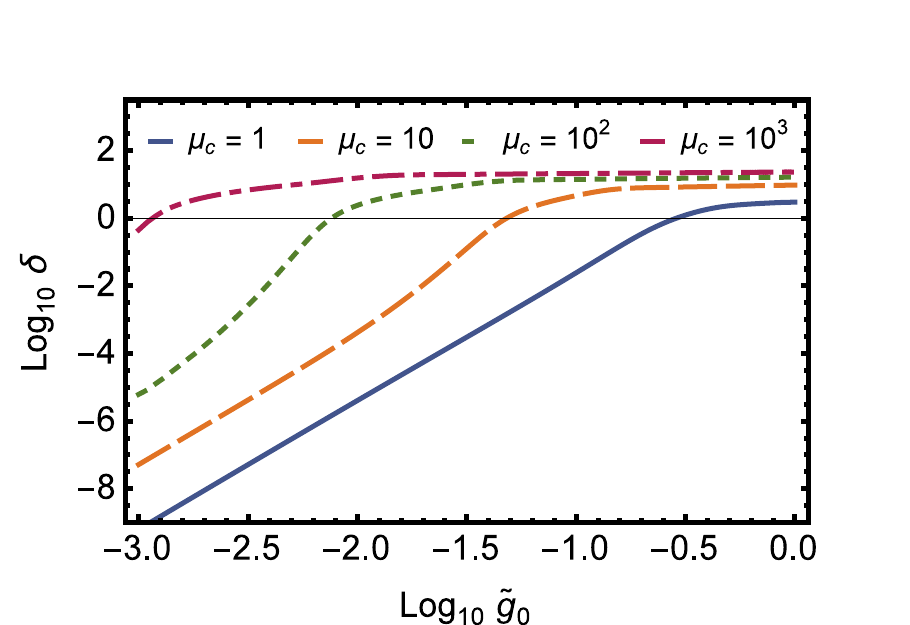}%
} \hfill
\subfloat[
\label{fig:measure:vs:mu}]{%
  \includegraphics[width=.42\linewidth, trim = 14mm 0mm 10mm 12mm]{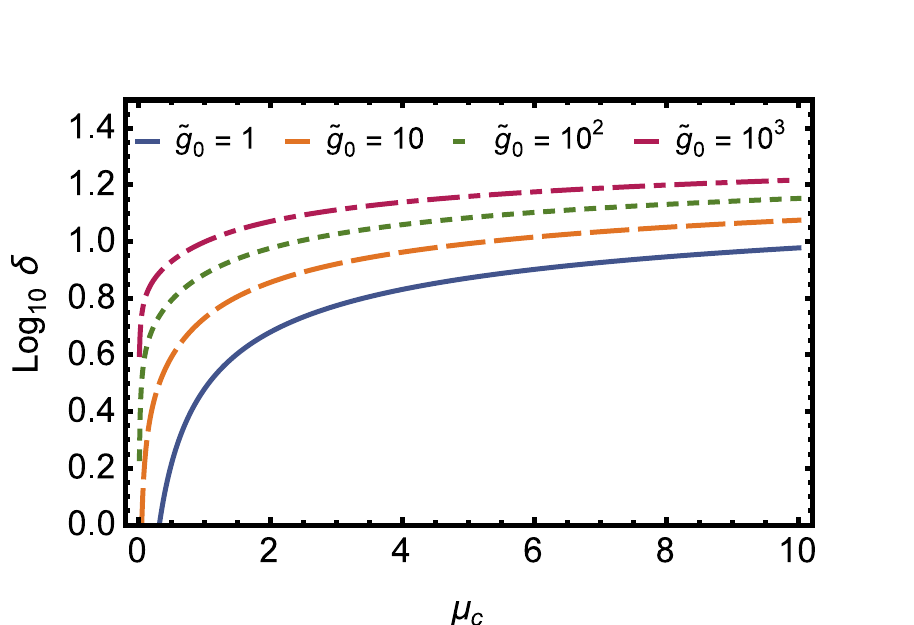}%
} \hfill
\caption{The behaviour of the measure of non-Gaussianity  $\delta(\tau)$ at $\tau = \pi$ for systems with constant nonlinear coupling $\tilde{g}_0$ starting in coherent states. \textbf{(a)}  A log--log plot of $\delta(\tau)$ vs. the rescaled coupling $\tilde{g}_0$. As $\tilde{g}_0$ increases, the state becomes more and more non-Gaussian, polynomially at first but then it quickly tends towards a constant value. \textbf{(b)} A log-plot of $\delta(\tau)$ vs. the coherent state parameter $\mu_{\mathrm{c}}$ for different values of $\tilde{g}_0$. $\delta(\tau)$ first increases quickly, then plateaus towards a single value. }
\label{fig:constant:measure:scaling}
\end{figure}

We now proceed to compute the exact measure of non-Gaussianity $\delta(\tau)$ for constant coupling $\tilde{g}_0$ and with the system initially in two coherent states. The exact expression is again too long and cumbersome to be reprinted here, but we plot the results in Figure~\ref{fig:constant:measure:vs:time} and Figure~\ref{fig:constant:measure:scaling}. In Figure \autoref{fig:measure:vs:time:various:mu} we plot the measure of non-Gaussianity $\delta(\tau)$ as a function of time $\tau$ for different values of the coherent state parameter $\mu_{\mathrm{c}}$, over the period $0<\tau<2\,\pi$. The other parameters are set to $\tilde{g}_0 = 1$ and $\mu_{\mathrm{m}} = 0$.  It is known that the full nonlinear dynamics is periodic (or recurrent) with period $2\,\pi$, see~\cite{qvarfort2018gravimetry}, whenever $\tilde{g}^2_0$ is an integer and this is clearly reflected in our plot.  

At $\tau = 2\pi$, the optics and mechanics are no longer entangled, and while the mechanics returns to its initial coherent or coherent thermal state (see Supplemental Note 1 in~\cite{qvarfort2018gravimetry} for an explicit proof), the final optical state will depends on the value of $\tilde{g}_0$. For example, when $\tilde{g}_0  = 0.5$, the cavity state becomes a superposition of coherent states at $\tau = 2\pi$, also known as a cat state~\cite{bose1997preparation}. However, if $\tilde{g}_0^2$ is integer, we obtain a phase factor of $e^{2\pi i \tilde{g}_0^2} = 1$ in the optical state, and the optics returns to an initial state as well. This is the case in Figure~\ref{fig:constant:measure:vs:time},  where $\delta(2\pi) = 0$. We will make use of the asymptotic measure defined in Section~\ref{sec:asymptotic:large:coherent:state} to analyse this behaviour, see Section~\ref{subsec:large:coherent:state}. Furthermore, while it might seem that the non-Gaussianity peaks at $\tau = \pi$, the measure $\delta$ exhibits a local minimum which grows increasingly narrow with larger $\mu_{\mathrm{c}}$. This is apparent from Figure~\ref{fig:measure:vs:time:various:mu:around:pi} where we have shown a close-up of $\delta$ around $\tau = \pi$ for increasing values of $\mu_{\mathrm{c}}$, and for $\tilde{g}_0 = 1$ and $\mu_{\mathrm{m}} = 0$.  The dip occurs because at $\tau = \pi$, we find that $\theta_a = - 2 \pi \tilde{g}^2_0$ and $F = - 2 \tilde{g}_0$. 
Thus, for integer $\tilde{g}_0^2$, we have $\sin^2{\theta_a/2} = 0$ and $\sigma_{11}$ becomes $\sigma_{11} = 1 + 2 |\mu_{\mathrm{c}}|^2 \left(1 - e^{- 4 \tilde{g}_0^2} \right)$. The non-zero exponent causes the non-Gaussianity to temporarily decrease, and the same behaviour occurs in the other covariance matrix elements, resulting in the dip.

As already noted, increasing $\mu_{\mathrm{c}}$ yields a logarithmic increase in $\delta(\tau)$, which is evident from the approximation in Eq.~\eqref{eq:delta:large}. 
Figure~\ref{fig:measure:vs:time:various:mu} also implies that for closed dynamics, the nonlinear system will almost immediately become maximally non-Gaussian. It will then retain approximately the same amount of non-Gaussianity until $\tau = 2\pi$, meaning there will be a rapid decrease of non-Gaussianity before the system revives again. 
The appearance of these plateaus shows that the maximum amount of non-Gaussianity available during one cycle can be accessed almost immediately without requiring the system to evolve for a long time. As a side remark, we note that the functional form of $\delta(\tau)$ in Figure~\ref{fig:measure:vs:time:various:mu}  closely resembles the linear entropy of the traced-out subsystems as found in ~\cite{bose1997preparation, qvarfort2018gravimetry}.

To better understand the behaviour of $\delta(\tau)$ for small times $\tau$,  we plot the behaviour of $\log_{10} {\delta(\tau)}$ for $\tau \ll 1$ for different values of $\mu_{\mathrm{c}}$ in Figure~\ref{fig:measure:for:small:tau}. We note that $\delta(\tau)$ increases quickly at first, but soon tends to a near-constant value. This means that $\delta(\tau)$ grows linearly for an interval of small times, which can be seen as the increasing and decreasing parts in Figure~\ref{fig:measure:vs:time:various:mu}. 

Finally, we proceed to examine the scaling behaviour of $\delta(\tau)$ at fixed time $\tau = \pi$. Figure~\ref{fig:measure:vs:C1} shows a $\log_{10}$--$\log_{10}$ plot of the measure $\delta(\tau)$ as a function of the nonlinear coupling $\tilde{g}_0$ for different values of $\mu_{\mathrm{c}}$. As $\tilde{g}_0$ increases, the amount of non-Gaussianity first grows linearly in the logarithm, then plateaus as $\tilde{g}_0$ increases further. The same behaviour occurs for larger $\mu_{\mathrm{c}}$, only more rapidly. This suggests that if we wish to increase the non-Gaussianity substantially, it will become increasingly difficult to do so by increasing $\tilde{g}_0$. As such, focusing on increasing the coupling $\tilde{g}_0$ will only give marginal returns. Similarly,~\ref{fig:measure:vs:mu} shows $\log_{10}{\delta(\tau)}$ as a function of increasing $\mu_{\mathrm{c}}$ for various values of $\tilde{g}_0$.

\subsection{Small coherent state parameters}
For a small amplitude coherent state of the optics, with $|\mu_{\mathrm{c}}|^2 \ll 1$, and with the mechanics in a coherent state, we found in~\eqref{eq:delta:small} that $\delta(\tau)$ scales with $\sim |F|^2 |\mu_{\mathrm{c}}|^2 \ln |\mu_{\mathrm{c}}|$. Given the explicit form of $F$, we see that it scales with $F\propto \tilde{g}_0$. Since $\delta(\tau)$ in this regime is proportional to $|F|^2$, it follows that $\delta(\tau)$ grows quadratically with the light--matter coupling in this regime.

\subsection{Large coherent state parameters} \label{subsec:large:coherent:state}

We derived an asymptotic form of $\delta(\tau)$  in~\eqref{eq:delta:large} for the case $|\mu_{\mathrm{c}}|\gg 1$, which we called $\tilde{\delta}(\tau)$. As argued before, the behaviour of the measure $\delta(\tau)$ in this regime depends crucially on the distance of $\theta_a$ from the value $2\pi$. In our present case we have that  $\theta_a\sim\tau^3$ for $\tau\ll1$ and $\theta_a\sim-\tilde{g}_0^2\tau$ for $\tau \gg1$. The functions that we decided to ignore (except for $\sigma_{11}$) in the derivation of $\tilde{\delta}(\tau)$ are of the form $f_{|\mu_{\mathrm{c}}|}(\theta)=(1-\exp[-\beta\,|\mu_{\mathrm{c}}|^2\sin(\theta_a/2)])$ or  $f_{|\mu_{\mathrm{c}}|}(\theta)=(1-\exp[-\beta'\,|\mu_{\mathrm{c}}|^2\sin^2(\theta_a/2)])$ where $\beta$ and $\beta'$ are irrelevant numerical constant of order $1$. We focus on $f_{|\mu_{\mathrm{c}}|}(\theta)$ and note that a similar argument applies for the other function as well. Finally, we ignore the transient regime of $\tau\ll1$ and focus on times $\theta_a\sim-\tilde{g}_0^2\theta$.

\begin{figure}[t!]
\subfloat[ \label{fig:approximate:measure:vs:time:various:mu}]{%
  \includegraphics[width=.42\linewidth, trim = 14mm 0mm 10mm 12mm]{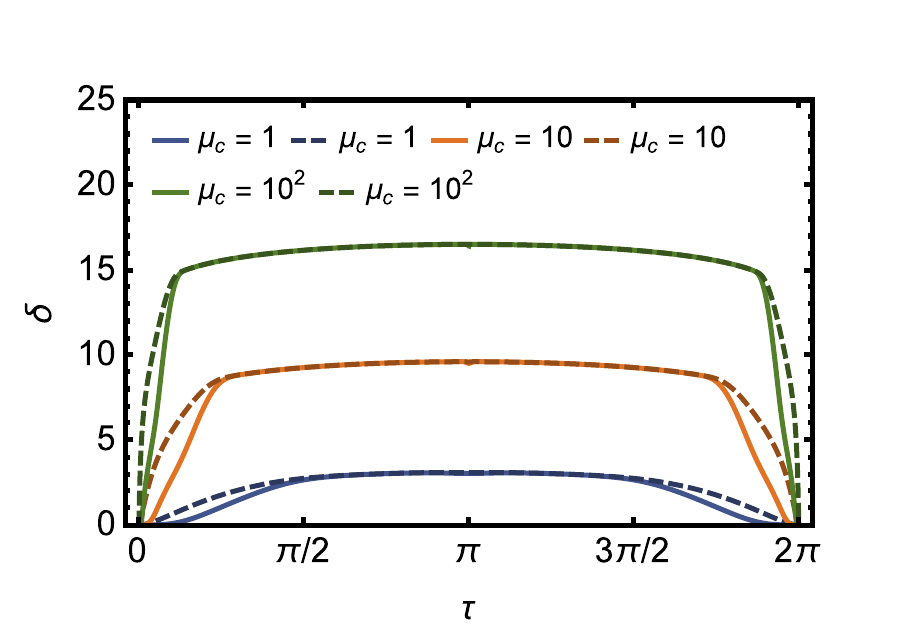}%
}\hfill
\subfloat[
\label{fig:approximate:measure:vs:time:various:g0}]{%
  \includegraphics[width=.42\linewidth, trim = 14mm 0mm 10mm 12mm]{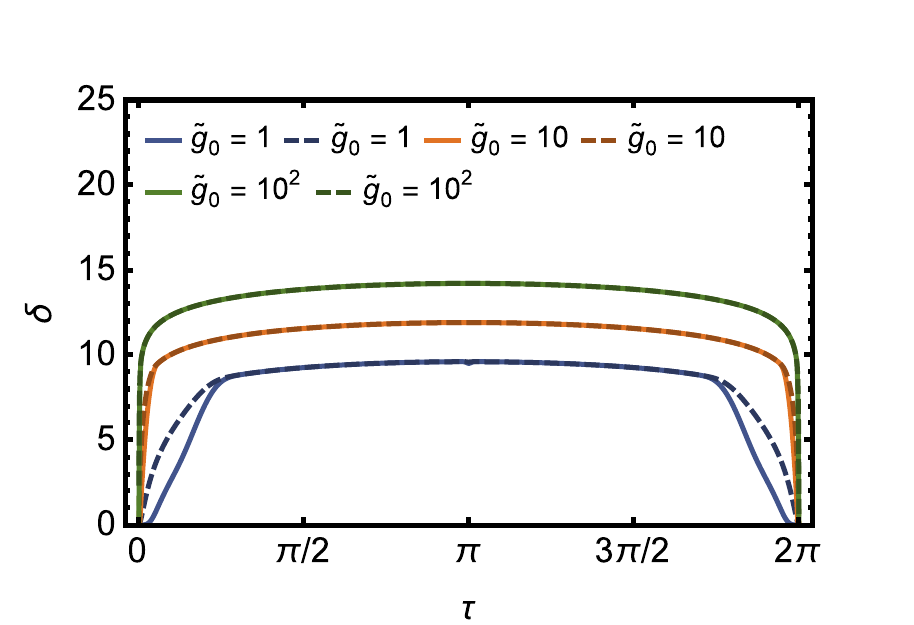}%
}
\caption{ Comparing the measure of non-Gaussianity $\delta(\tau)$ (solid lines) with the asymptotic form computed in Eq.~\eqref{eq:delta:large} (dashed lines) for coherent states. \textbf{(a)} Exact measure (solid line) vs. the approximation for different values of $\mu_{\mathrm{c}}$. As $\mu_{\mathrm{c}}$ increases, the approximation grows increasingly accurate. In this plot, $\tilde{g}_0 = 1$. \textbf{(b)} Exact measure (solid line) vs. the approximation for large $\mu_{\mathrm{c}}$ for increasing values of $\tilde{g}_0$ and $\mu_{\mathrm{c}} = 10$. The approximation becomes increasingly accurate as $\tilde{g}_0$ increases, even towards the beginning and end of one oscillation period. }
\label{fig:measure:compared:with:approximation}
\end{figure}

To see how well the asymptotic form $\tilde{\delta}(\tau)$ in ~\eqref{eq:delta:large} approximates the exact measure, we have plotted both the exact form of $\delta(\tau)$ (solid lines) with the asymptotic form (dashed lines) in Figure~\ref{fig:measure:compared:with:approximation}. We note that, even for $|\mu_{\mathrm{c}}|\sim 1$, the asymptotic measure $\tilde{\delta}(\tau)$ well approximates the exact value of $\delta(\tau)$. In fact, it becomes even more accurate as the optical coherent state parameter $\mu_{\mathrm{c}}$ increases, which is to be expected given the nature of the approximation. The asymptotic form also becomes  more accurate once we also increase $\tilde{g}_0$, as evident in Figure~\ref{fig:approximate:measure:vs:time:various:g0}. For  $\tilde{g}_0 = 10^2$, the approximation is almost entirely accurate. This occurs because the function $\theta_a$ increases with $\tilde{g}_0$, which further suppresses the off-diagonal covariance matrix elements at the beginning and end of each cycle.

Let us discuss the fact that the measure recurs with $\tau = 2\pi$ for integer $\tilde{g}_0^2$ which we can now address analytically by examining the asymptotic covariance matrix elements in~\eqref{covariance:matrix:elements:large:input:photons:far:from:critical:times}. We find that $F = 0$ for all $\tau = 2\pi n $ with integer $n$. This means that $\sigma_{42} = 0$ and that $\sigma_{22} = 1$. We also find that $\theta_n(2\pi)  = - 4 \pi \tilde{g}_0^2 $. Thus, if $\tilde{g}_0^2$ is integer, we find that $\sin^2{\theta_a/2} = 0$ and the final covariance matrix element is $\sigma_{11} = 1$. This results in $\mathbf{\sigma} = \mathrm{diag}(1,1,1,1)$ which corresponds to a coherent state, which is fully Gaussian. As a result, the non-Gaussianity vanishes. When $\tilde{g}_0^2$ is not an integer, some non-Gaussianity will be retained, but the fact that $F = 0$ will still result in a reduction at $\tau = 2\pi$. 

\subsection{Non-Gaussianity in open systems with constant coupling} \label{subsec:open:system:constant:coupling}
\begin{figure}[t!]
\subfloat[ \label{fig:open:system:photon:decoherence:constant:coupling}]{%
  \includegraphics[width=.42\linewidth, trim = 14mm 0mm 10mm 5mm]{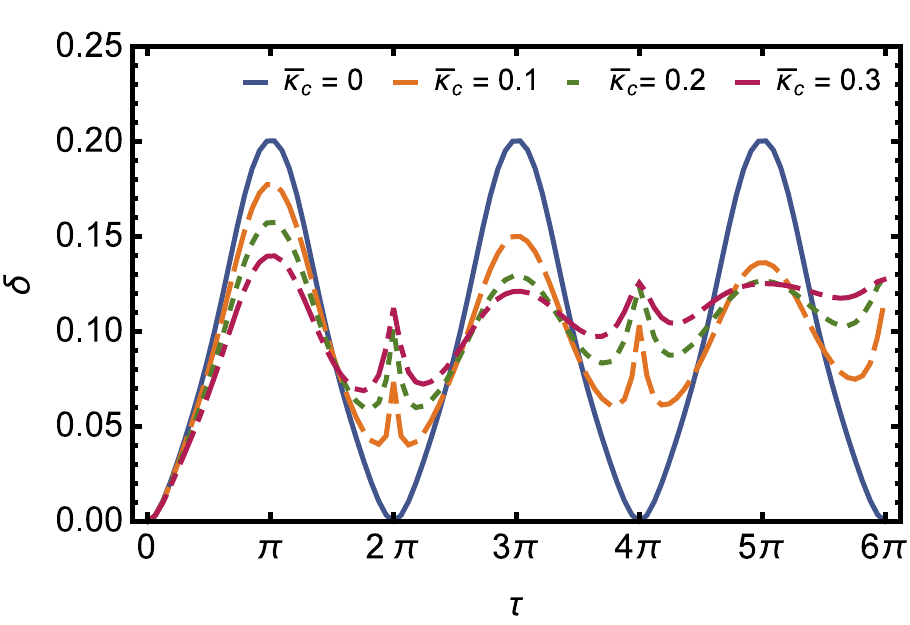}%
}\hfill
\subfloat[
\label{fig:open:system:phonon:decoherence:constant:coupling}]{%
  \includegraphics[width=.42\linewidth, trim = 14mm 0mm 10mm 5mm]{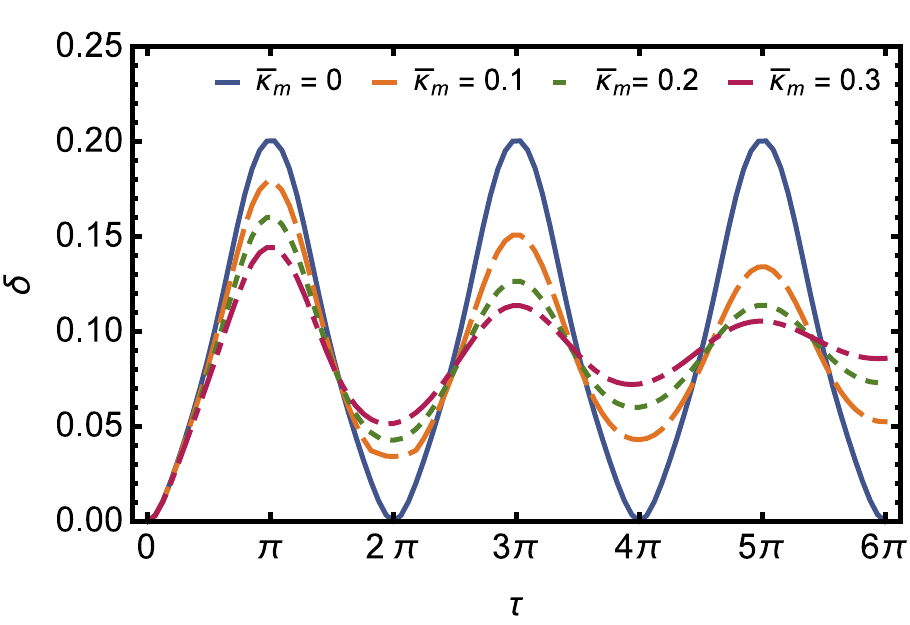}%
}
\caption{Non-Gaussianity of open optomechanical systems with constant light--matter coupling starting in a coherent state. \textbf{(a)} Non-Gaussianity $\delta $ vs. time $\tau$ for a system with increasing values of photon decoherence $\bar{\kappa}_{\mathrm{c}}$ for $\tilde{g}_0 = 1$, $\mu_{\mathrm{c}} = 0.1$ and $\mu_{\mathrm{m}} = 0$. \textbf{(b)} Non-Gaussianity $\delta$ vs. time $\tau$ for a system with increasing values of phonon decoherence $\bar{\kappa}_{\mathrm{m}}$ for $\tilde{g}_0 = 1$, $\mu_{\mathrm{c}} = 0.1$ and $\mu_{\mathrm{m}} = 0$. A populated mechanical coherent state $\mu_{\mathrm{m}} \neq 0$ does not affect the non-Gaussianity.}
\label{fig:measure:decoherence:constant:coupling}
\end{figure}

Any realistic system will suffer from decoherence. In Figure~\ref{fig:measure:decoherence:constant:coupling} we have plotted the non-Guassianity $\delta(\tau)$ as a function of time for an optomechanical system with open dynamics. Here, the cavity state and the mechanics are both in initial coherent states ~\eqref{initial:state:two}. Figure~\ref{fig:open:system:photon:decoherence:constant:coupling} shows the non-Gaussianity for increasing values of the photon decoherence rate $\bar{\kappa}_{\mathrm{c}} = \kappa_{\mathrm{c}}/\omega_{\mathrm{m}}$ with Lindblad operator $\hat L _{\mathrm{c}} = \sqrt{\bar{\kappa}_{\mathrm{c}}} \, \hat a$ and values $\mu_{\mathrm{c}} = 0.1$, $\tilde{g}_0  = 1$ and $\mu_{\mathrm{m}} = 0$. We have chosen a low value of $\mu_{\mathrm{c}} $ to ensure high numerical accuracy of the simulation, as larger values quickly lead to numerical instabilities. We note that the non-Gaussianity $\delta(\tau)$ tends towards a steady value, which is clear from the fact that the higher values of decoherence start to coincide around $\tau = 5\pi$. We also note that around $\tau = 2\pi n$, for integer $n$ the inclusion of noise appears to temporarily increase the non-Gaussianity. This could, however, be due to the fact that the relative entropy measure cannot distinguish between non-Gaussianity induced as a result of genuinely nonlinear dynamics or as a result of classical mixing of the states~\cite{barbieri2010non}. We discuss this further in Section~\ref{sec:discussion}. Similarly, in Figure~\ref{fig:open:system:phonon:decoherence:constant:coupling} we have plotted the non-Gaussianity $\delta(\tau)$ for increasing values of phonon decoherence rate $\bar{\kappa}_{\mathrm{m}}$ with Lindblad operator $\hat L _{\mathrm{m}} = \sqrt{\bar{\kappa}_{\mathrm{m}}} \, \hat b$ and the same values as before.

\section{Applications: Time-dependent coupling} \label{sec:time:dependent:coupling}
In all physical systems, such as optomechanical cavities, the confining trap is not ideal. This means that, in general, the coupling $\tilde{g}(\tau)$ is time-dependent as a consequence of, for example, trap instabilities. Time-dependent variations such as phase fluctuations in the laser beam used to trap a levitated bead will modulate the coupling. 

In this work, we want to exploit the possibility of controlling the coupling $\tilde{g}(\tau)$ by considering its periodic modulation in time.
In practice, such time-dependent control would be achievable for an optically trapped and levitated dielectric bead that interacts with a cavity field by controlling the optical phase of the trapping laser field. In fact, such phase determines the bead's equilibrium position which, in turn, affects the cavity light-bead coupling through the varying overlap between the bead and the cavity mode function.
Technically, the trapping laser's optical phase may be controlled through an acoustic-optical modulator. 
Alternately, control on a bead's equilibrium position may also be enacted by adopting Paul traps, which work for levitated nanospheres~\cite{millen2015cavity}, and have been used to shuttle ions across large distances, typically for the purpose of quantum information processing~\cite{hensinger2006t, walther2012controlling}. See also Section~\ref{sec:discussion} for additional methods by which the coupling can be modulated. 

\subsection{Modelling the trap modulation}\label{trap:instability:section}
We shall model a time-dependent light--matter coupling $\tilde{g}(\tau)$ by assuming that the coupling has the simple form 
\begin{align}
\tilde{g}(\tau) =\tilde{g}_0\,\left( 1+ \epsilon\sin(\Omega_0 \,\tau) \right).
\end{align}
Here, $\tilde{g}_0$ is the expected value of the coupling, $\epsilon$ is the amplitude of oscillation and $\Omega_0:=\omega_0/\omega_\textrm{m}$ is the dimensionless frequency that determines how the coupling oscillates in time. We can insert this ansatz in the general expressions~\eqref{sub:algebra:decoupling:solution:text} and obtain an explicit form for this case. The full expressions for the coefficients in~\eqref{sub:algebra:decoupling:solution:text} are again very long and cumbersome, and we do not print them here. They are listed in~\eqref{eq:time:dependent:coefficients}. 

\begin{figure}[t!]
\centering
\subfloat[ \label{fig:time:dependent:plain}]{%
  \includegraphics[width=.42\linewidth, trim = 10mm 0mm 10mm 12mm]{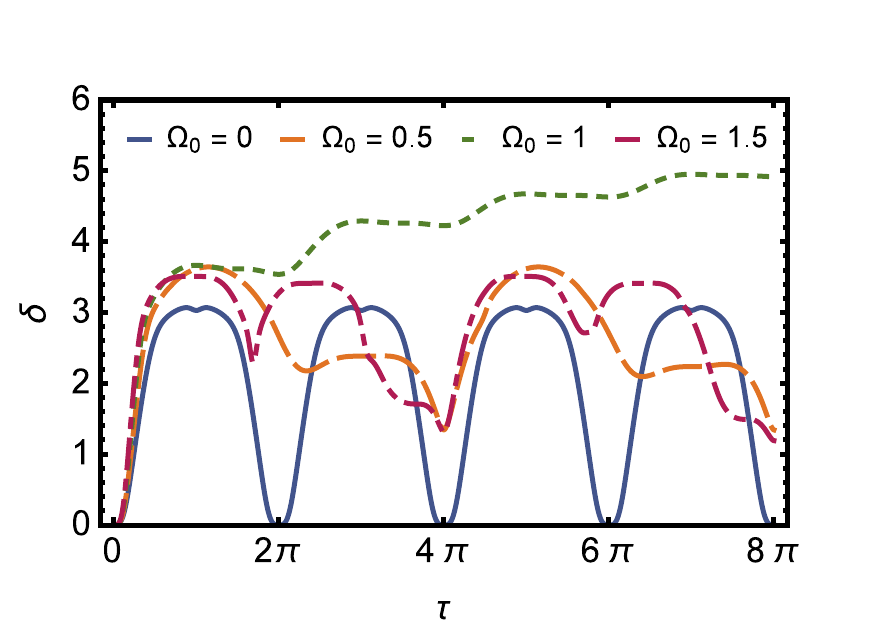}%
}\hfill
\subfloat[
\label{fig:resonance}]{%
  \includegraphics[width=.42\linewidth, trim = 10mm 0mm 10mm 12mm]{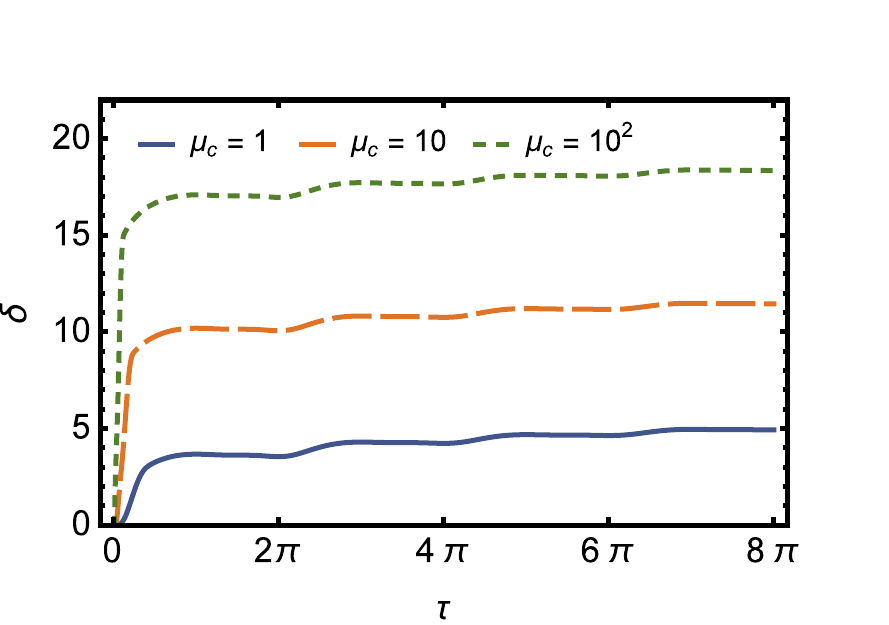}%
}\hfill
\subfloat[
\label{fig:resonance:two}]{%
  \includegraphics[width=.42\linewidth, trim = 10mm 0mm 10mm 12mm]{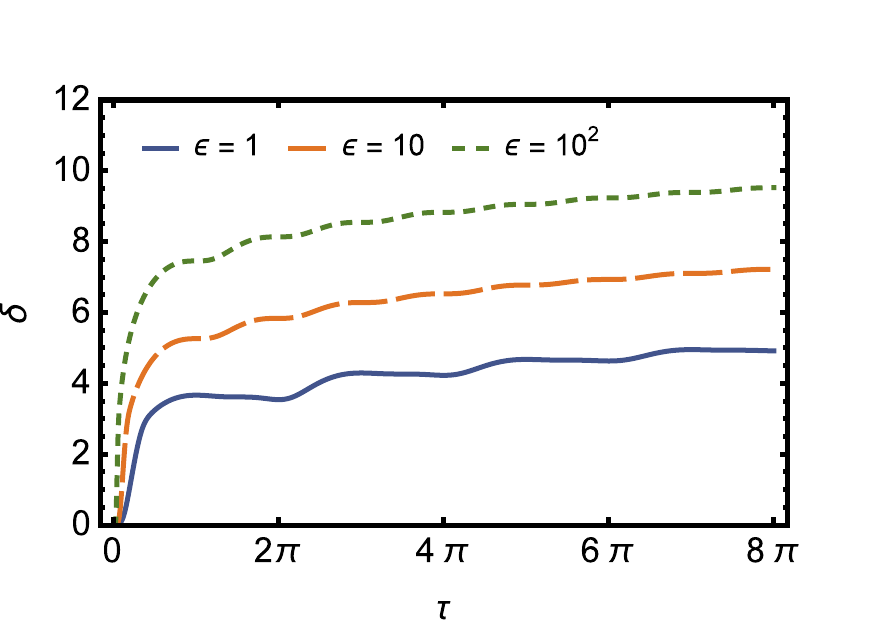}%
}
\caption{The measure of non-Gaussianity  $\delta(\tau)$ for systems with time-dependent coupling $\tilde{g}(\tau) = \tilde{g}_0(1 + \epsilon \sin{(\Omega_0 \tau)} )$, where $\epsilon$ is the amplitude and $\Omega_0 = \omega_0/\omega_{\mathrm{m}}$ is the modulation frequency. \textbf{(a)} Plot of $\delta(\tau)$ vs. rescaled time $\tau$ for different values of $\Omega_0$. The case $\Omega_0 = 0$ (blue line) corresponds to the time-independent setting. At resonance, with $\Omega_0 = 1 $ (green line), the system displays a drastically different behaviour. Other parameters are $\tilde{g}_0 = \epsilon = 1$ and $\mu_{\mathrm{m}} = 0$. \textbf{(b)} Plot of $\delta(\tau)$ vs. rescaled time $\tau$ at resonance $\Omega_0 = 1$ for various values of coherent state parameter $\mu_{\mathrm{c}}$. The system no longer exhibits closed dynamics. Other parameters include $\tilde{g}_0 = \epsilon = 1$ and $\mu_{\mathrm{m}} = 0$. \textbf{(c)} A plot of $\delta(\tau)$ vs. time $\tau$ for increasing oscillation frequency $\epsilon$ at resonance $\Omega_0 = 1$ and with $\mu_{\mathrm{c}}  = 1$. $\delta(\tau)$ increases slowly with $\epsilon$. Again, we have set $\mu_{\mathrm{m}} = 0$. }
\label{fig:time:dependent}
\end{figure}

We can now compute $\delta(\tau)$ for this time-dependent coupling for initial coherent states, and we display the results in Figure~\ref{fig:time:dependent}. In~\ref{fig:time:dependent:plain} we plot $\delta(\tau)$ vs. $\tau$ for different values of the oscillation frequency $\Omega_0$. Note that we here include a larger range of $\tau$ to capture potentially recurring behaviour. In the limit $\Omega_0 \rightarrow 0$ we recover the time-independent solution, as expected. Interestingly, when $\Omega_0\neq0$ we see that we can achieve higher values for the nonlinear measure $\delta(\tau)$. This is especially pronounced as $\Omega_0\rightarrow 1$, where the trap oscillation frequency is equal to the mechanical frequency $\omega_{\mathrm{m}}$, for which $\delta(\tau)$ ceases to oscillate periodically, but instead steadily increases. We discuss this case in detail in the following section.

\subsection{Trap modulation on resonance}\label{resonance}
The functions~\eqref{eq:time:dependent:coefficients} contain denominators of the form $\Omega_0-1$. Therefore, among all possible values of $\Omega_0$, we can ask what happens \textit{on resonance}, i.e., when $\Omega_0=1$. Figure~\ref{fig:time:dependent:plain} already provides evidence that the system should behave markedly differently. 

At resonance, where $\Omega_0 = 1$, the functions~\eqref{eq:time:dependent:coefficients} take the relatively simple form
\begin{align}\label{eq:resonance:coefficients}
F_{\hat N_a^2} &=-\frac{1}{16} \tilde{g}_0 \, \bigl[ 16 \, \tau-8 \sin (2\, \tau)+\epsilon \, (32-36 \cos (\tau)+4 \cos (3 \, \tau))\nonumber \\
&\quad\quad\quad\quad\quad\quad\quad +\epsilon ^2 \,   \bigl( 6 \, \tau-4 \sin (2\, \tau)+\sin (2 \, \tau)\,\cos (2 \, \tau) \bigr)\bigr]\nonumber\\
F_{\hat N_a \, \hat B_+} &= -\tilde{g}_0 \sin (\tau) \left(1+ \frac{\epsilon}{2} \sin (\tau)\right) \nonumber \\
F_{\hat N_a \, \hat B_-} &=\frac{\tilde{g}_0}{4} \epsilon   \,  \left(  \sin (2 \, \tau)-2 \, \tau  \right) -  2 \, \tilde{g}_0 \,  \sin^2\left( \frac{\tau}{2} \right)
\end{align}
We have plotted in Figure~\ref{fig:time:dependent} the exact measure of non-Gaussianity $\delta(\tau)$ in the resonant case for initially coherent states and for different values of $\mu_{\mathrm{c}}$. As anticipated, here we no longer have recurrent behaviour. Instead, the non-linearity increases as $\ln{\tau}$. Formally, this growth can continue for arbitrarily large times $\tau$, however, the maximum time $\tau$ that can be achieved in practice is limited by the coherence time of the experiment. Similarly, we plotted $\delta(\tau)$ for various values of $\epsilon $ in Figure~\ref{fig:resonance:two}. We note that $\delta(\tau)$ oscillates increasingly rapidly with larger $\epsilon$ but with decreasing amplitude for increasing $\tau$, as $|F|^2\sim\tilde{g}_0^2\,\epsilon^2\,\tau^2$ becomes the dominant term for $\tau \gg 1$.

As already noted, it is evident from Figure~\ref{fig:time:dependent} that the non-Gaussianity increases continuously. 
The nonlinear coupling in the Hamiltonian is derived by considering the effect of photon pressure on the mechanical element. Given that the overall photon number $\langle \hat a ^\dag \hat a \rangle $ is conserved, the coupling acts as a photon number displacement. If this coupling is time-dependent, this means that the photon pressure displaces with a time-dependence. When this occurs at resonance, this linear displacement grows linearly in time. See also~\cite{liao2014modulated} for further insight once the rotating wave approximation has been applied. 

\subsection{Large coherent state parameters at resonance}\label{subsec:large:coherent:state:resonance}
Using the explicit form of the coefficients~\eqref{eq:resonance:coefficients}, we note that $|F|^2 = F_{\hat N_a \, \hat B_-}^2 + F_{\hat N_a \, \hat{B}_+}^2$ has the asymptotic behaviour $|F|^2\sim\frac{1}{4}\tilde{g}_0^2\,\epsilon^2\,\tau^2$ for $\tau\gg1$. This implies that $\exp[-|F|^2]\ll1$ for large $\tau$ and therefore we expect, as it happened in Section~\ref{sec:asymptotic:large:coherent:state}, that most covariance matrix elements  will vanish and will not contribute to the asymptotic form of $\delta(\tau)$. This observation allows us to compute the symplectic eigenvalues, which  read $\nu_+=1 + 2|\mu_{\mathrm{c}}|^2 \left( 1 - e^{-4 \, |\mu_{\mathrm{c}}|^2 \sin^2{\theta_a/2}} \, e^{- |F|^2}  \right)$ and $\nu_-=\sqrt{1+4\,|\mu_{\mathrm{c}}|^2\,|F|^2}$, and they 
match the expressions~\eqref{eq:approx:symplectic:eigenvalues}. We again stress that we have retained the exact expression for $\sigma_{11}$ to capture some crucial features of the non-Gaussianity, such as $\delta(0) = 0$. 

\begin{figure}[t!]
\subfloat[ \label{fig:resonance:comparison:various:mu}]{%
  \includegraphics[width=.42\linewidth, trim = 10mm 0mm 10mm 12mm]{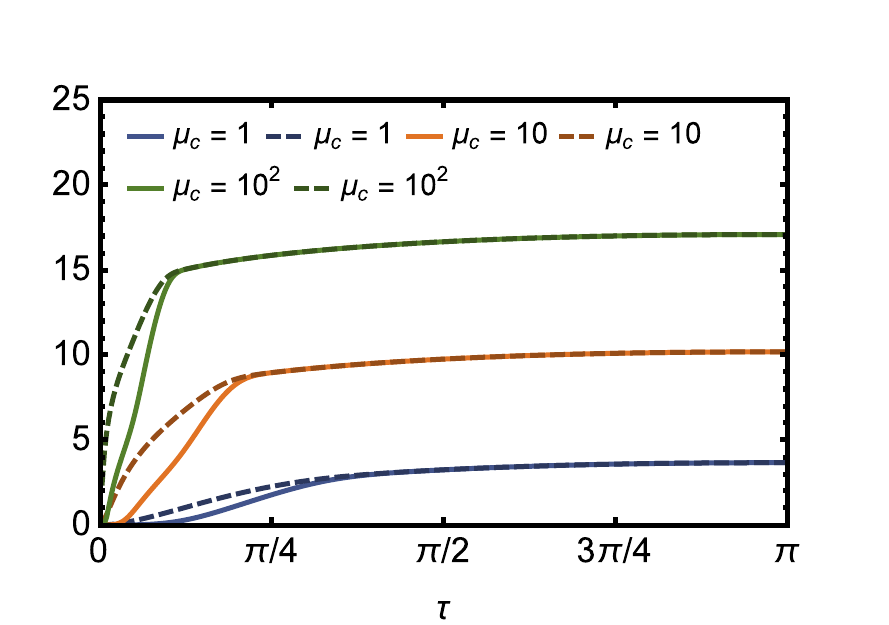}%
}\hfill
\subfloat[
\label{fig:resonance:comparison:various:delta:g}]{%
  \includegraphics[width=.42\linewidth, trim = 10mm 0mm 10mm 12mm]{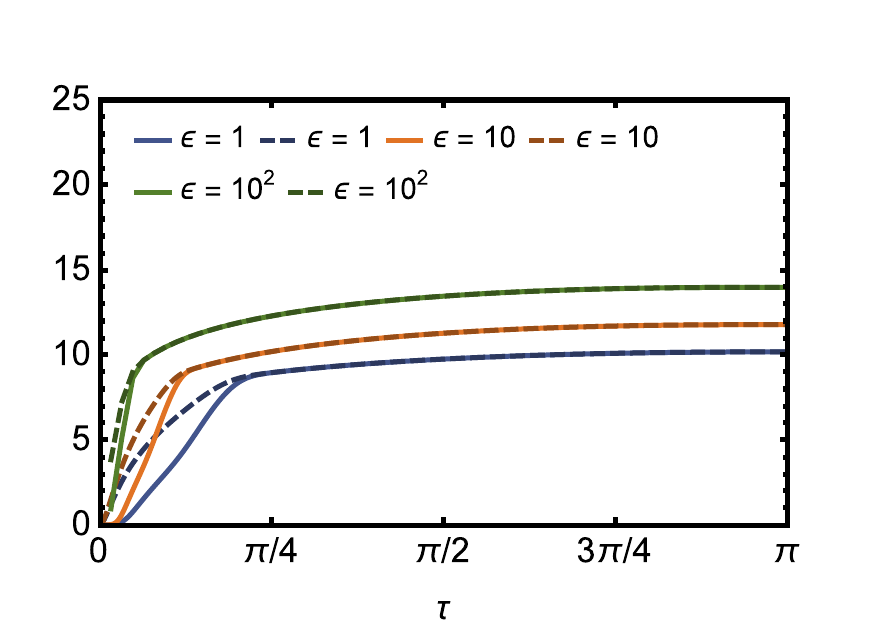}%
}
\caption{A comparison between the full measure $\delta(\tau)$ (solid line) and the approximate measure (dashed lines) for time-dependent couplings $\tilde{g}(\tau)$. \textbf{(a)} Plot showing the accuracy of the approximation for different values of $\mu_{\mathrm{c}}$ at $\Omega_0 = 0.5$. The approximation becomes very accurate as $\mu_{\mathrm{c}}$ increases. \textbf{(b)} Plot comparing the accuracy of $\tilde{\delta}$ for a different values of $\epsilon$ at $\mu_{\mathrm{c}} = 10$. The approximation becomes more accurate as $\epsilon$ increases. }\label{fig:time:dependent:comparison}
\end{figure}

In Figure~\ref{fig:time:dependent:comparison}, we compare the exact measure $\delta(\tau)$ at resonance with the asymptotic form derived  in~\eqref{eq:delta:large}. The solid lines represent the exact measure $\delta(\tau)$ and the dashed lines represent the asymptotic expression. In Figure~\ref{fig:resonance:comparison:various:mu} we compare them for different values of $\mu_{\mathrm{c}}$. We note that, except for at very small $\tau$, the asymptotic form is entirely accurate and gets even more precise for increasing values of $\mu_{\mathrm{c}}$. This is a consequence, as we noted before, of the exponentials in~\eqref{full:elements:covaraince:matrix} that suppress some elements for large $\mu_{\mathrm{c}}$, unless $\theta_a=n\,\pi$. Similarly, in Figure~\ref{fig:resonance:comparison:various:delta:g} we have plotted $\delta(\tau)$ and its asymptotic form for different values of the oscillation amplitude $\epsilon$. Again, the suppression of the exponentials with increasing $\epsilon$ means that larger values of $\epsilon$ yield a more accurate expression.

\subsection{Open system dynamics at resonance}\label{subsec:open:system:resonance}
\begin{figure}[t!]
\subfloat[ \label{fig:open:system:photon:decoherence:resonance}]{%
  \includegraphics[width=.42\linewidth, trim = 14mm 0mm 10mm 5mm]{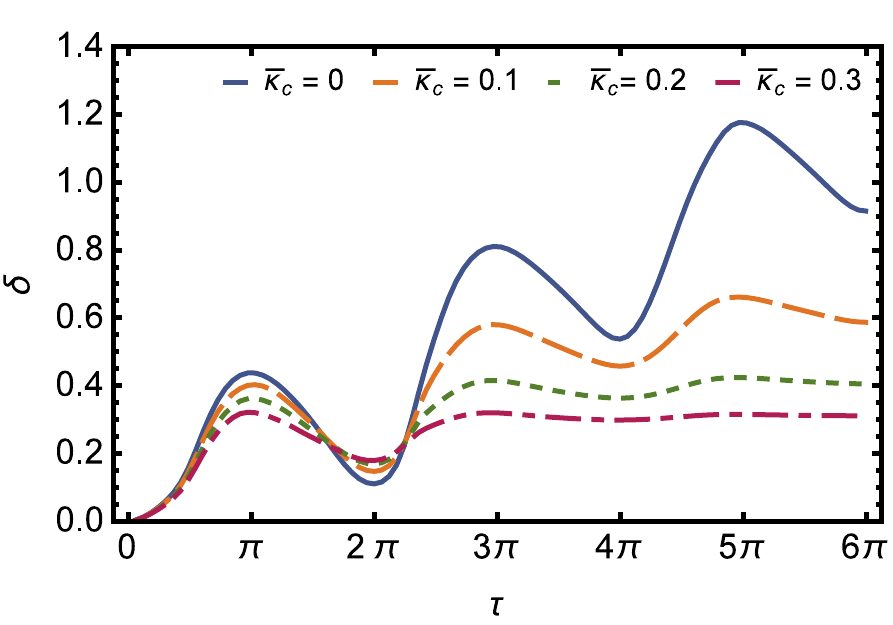}%
}\hfill
\subfloat[
\label{fig:open:system:phonon:decoherence:resonance}]{%
  \includegraphics[width=.42\linewidth, trim = 14mm 0mm 10mm 5mm]{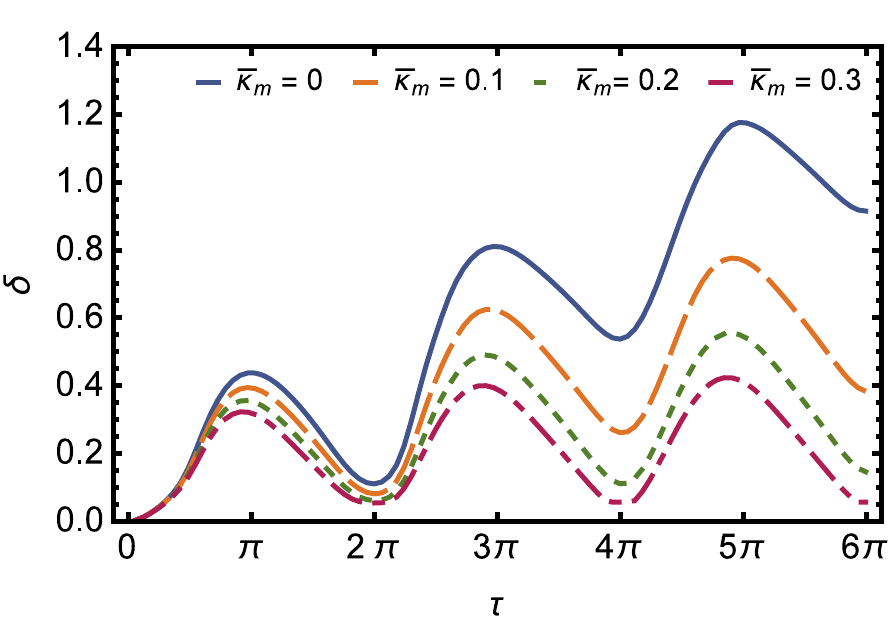}%
}
\caption{Non-Gaussianity for open optomechanical systems at mechanical resonance. \textbf{(a)} Non-Gaussianity $\delta $ vs. time $\tau$ for a system with increasing values of photon decoherence $\bar{\kappa}_{\mathrm{c}}$ for $\tilde{g}_0 = 1$, $\mu_{\mathrm{c}} = 0.1$, $\epsilon = 0.5$,  and $\mu_{\mathrm{m}} = 0$. \textbf{(b)} Non-Gaussianity $\delta$ vs. time $\tau$ for a system with increasing values of phonon decoherence $\bar{\kappa}_{\mathrm{m}}$ for $\tilde{g}_0 = 1$, $\mu_{\mathrm{c}} = 0.1$, $\epsilon = 0.5$,  and $\mu_{\mathrm{m}} = 0$. Changing to $\mu_{\mathrm{m}} \neq 0 $ does not affect the results. }
\label{fig:measure:decoherence:resonance}
\end{figure}

If it is possible to continuously increase the non-Gaussianity, the system might have a certain tolerance to noise. That is, there is a level of noise at which the non-Gaussianity essentially reaches a steady-state. In Figure~\ref{fig:measure:decoherence:resonance} we have plotted the non-Gaussianity $\delta$ as a function of time for different values of photon and phonon decoherence. Figure~\ref{fig:open:system:photon:decoherence:resonance} shows the system at resonance with photons leaking from the cavity with a rate $\bar{\kappa}_{\mathrm{c}} = \kappa_{\mathrm{c}} / \omega_{\mathrm{m}}$ for parameters $\mu_{\mathrm{c}} = 0.1$, $\tilde{g}_0 = 1$, $\epsilon = 0.5$ and $\mu_{\mathrm{m}} = 0$. We note that $\bar{\kappa}_{\mathrm{c}} = 0.3$ yields what is essentially a steady-state of the non-Gaussianity. In Figure~\ref{fig:open:system:phonon:decoherence:resonance} we note the same behaviour but for phonon decoherence with rate $\bar{\kappa}_{\mathrm{m}} = \kappa_{\mathrm{m}} / \omega_{\mathrm{m}}$.

\section{Discussion and practical implementations}\label{sec:discussion}
We have employed a measure of non-Gaussianity $\delta(\tau)$ in order to quantify the deviation from linearity of an initial Gaussian state induced by the Hamiltonian~\eqref{main:time:independent:Hamiltonian:to:decouple}. Our results show that, for a constant light--matter coupling $\tilde{g}_0$, the non-Gaussianity $\delta(\tau)$ scales differently in two contrasting regimes: (i)~For a weak optical input coherent state $\lvert\mu_{\mathrm{c}}\rvert$, the nonlinear character of the state grows as $\tilde{g}_0^2\lvert\mu_{\mathrm{c}}\lvert^2 \, \ln{\lvert\mu_{\mathrm{c}}\lvert}$ if the mechanics is also in a coherent state,  (ii)~conversely, for large $\lvert\mu_{\mathrm{c}}\lvert$, the nonlinear character of the state grows logarithmically with the quantity $\tilde{g}_0 |\mu_{\mathrm{c}}|$, which also holds when the mechanical element is not fully cooled. The same general scaling with $|\mu_{\mathrm{c}}|$ occurs when $\tilde{g}(\tau)$ is time-dependent.  

Crucially, we also find that the amount of non-Gaussianity can be continuously increased by driving the light--matter coupling at mechanical resonance. This becomes especially useful in the presence of noise. We will now discuss these results in the context of concrete experimental setups, and specifically discuss how the modulated light--matter coupling can be engineered. First, however, we will discuss the measure of non-Gaussianity that we have used in this work.

\subsection{Choice of measure}\label{subsec:measure:discussion}
In this work, we chose to work with a relative entropy measure of non-Gaussianity (see Section~\ref{sec:measure:non:gaussianity}) which was first defined in~\cite{genoni2008quantifying}. This measure has previously been extensively used to compute the non-Gaussianity of various states~\cite{tatham2012nonclassical}, as well as in an experimental setting where single photons were gradually added to a coherent state to increase its non-Gaussian character~\cite{barbieri2010non}. Several additional measures for the quantification of non-Gaussianity have been proposed in the literature, linking it to the Hilbert-Schmidt distance~\cite{genoni2007measure} or to quantum correlations~\cite{park2017quantifying}. Specifically, the relative entropy measure was shown to be more general than the Hilbert-Schmidt measure~\cite{genoni2007measure}. Furthermore, a connection has been put forward between the non-Gaussianity of a state and its Wigner function~\cite{genoni2013detecting}, and similarly there appears to be an intrinsic link between the quantum Fisher information and the lowest amount of non-Gaussianity of a state~\cite{yadin2018operational}.

Most crucially, this measure is not upper-bounded. This means that, as opposed to, for example, an entanglement measure where the notion of a maximally entangled state is well-defined, there is no such thing as a maximally non-Gaussian state. This is reflected by our results, where taking $\mu_{\mathrm{c}}$ to infinity yields $\lim_{\mu_{\mathrm{c}} \rightarrow \infty} \delta = \infty$. As such, it is only possible to state that one state is more non-Gaussian than another. However, for pure states, there is the relation of the measure to the Hilbert-Schmidt measure. As such, the non-Gaussianity $\delta(\tau)$ of pure states has strong operational implications~\cite{zhuang2018resource}.

For mixed states, the operational meaning is not clear because the measure cannot detect the difference between classical mixtures of Gaussian states, which can be easily prepared by classical mixtures of Gaussian states, and inherent non-Gaussianity due to some nonlinear evolution of pure states~\cite{barbieri2010non}. This means that the measure often needs to be used together with a measure of non-classicality, such as the negativity of the Wigner function. We know from previous work~\cite{bose1997preparation, qvarfort2018gravimetry} that for a constant coupling, the system is maximally entangled at $\tau = \pi$, which satisfies the occurrence of non-classicality in conjunction with the non-Gaussianity. The state is however fully disentangled at $\tau = 2\pi$, and in the case of open system dynamics, this feature of the measure becomes apparent. We note that the non-Gaussianity plotted in Figure~\ref{fig:measure:decoherence:constant:coupling}  spikes at times $\tau = 2\pi n$ for integer $n$, which is when we usually have no entanglement. This implies that the addition of non-Gaussianity most likely comes from a classical mixture of coherent states that have slightly decohered.

\subsection{Experimental regimes}
There are two relevant experimental regimes for optomechanical systems. They are determined by the magnitude of the light--matter coupling $g$ compared to the other frequencies in the system. In the weak single-photon optomechnical coupling regime, the light--matter coupling $g$ is small compared to the resonant frequency $\omega_{\mathrm{m}}$ and the optical decoherence rate $\kappa_{\mathrm{c}}$. Such experiments usually involve a strong laser drive, which tends to wash out the non-linearity. In the strong single-photon coupling regime, nonlinear effects are in practice small but more significant. Under these conditions, a single photon displaces the mechanical oscillator by more than its zero-point uncertainty and weak optical fields tend to be used~\cite{nunnenkamp2011single}. In summary, most approaches fall into one of two categories: (i) small $g$ and linearised dynamics and (ii) large $g$ and low number of photons. 

Our work suggests that we can further increase the amount of non-Gaussianity by modulating the light--matter coupling. We emphasize that this scheme is applicable in both the weak and strong coupling regimes. This sets it apart from other schemes, which usually focus on enhancing the non--Gaussianity in one of the two categories mentioned above. 

Let us also briefly discuss our results with regard to linearised dynamics. This linearisation of dynamics is fundamentally different to the scenarios considered in this work. When linearising the dynamics, the system is opened and the field operators $\hat a $ are treated as flucutations around a strong optical field as such: $\hat a \rightarrow \hat a = \alpha + \hat a'$, where $\hat a'$ are the fluctuations. In this work, we have retained the nonlinear dynamics, even when considering open system dynamics. Thus, while we observe that a large coherent state parameter $\mu_{\mathrm{c}}$ increases the non-Gaussianity, we cannot generalise this result to the linearised dynamics.

\subsection{Methods of modulating the light--matter coupling in physical systems}\label{subsec:modulating}
We saw in Section~\ref{sec:time:dependent:coupling} that the amount of non-Gaussianity in the system increases when the light--matter coupling $\tilde{g}(\tau)$ is modulated. An explanation of this phenomena was provided in~\cite{liao2014modulated}. Consider the force $\vec{F}$ exerted by the photons on the mechanics. For a number of $n$ photons, this force is proportional to $\vec{F} \propto (n + 1/2)$, where $1/2$ comes from the zero-point energy. When the light--matter coupling is constant, this force is constant, and thus we see the periodic evolution. However, when we modulate $\tilde{g}(\tau)$, the photon-pressure force $\vec{F}$ acts periodically on the mechanics, and is amplified when pushing in tandem with the mechanical resonance. 

While engineering the modulation is challenging, we shall explore several methods that can achieve it. The question is whether the modulation can be performed at mechanical resonance. As a basis for this discussion, we present a derivation of a time-dependent light--matter coupling for levitated nanobeads in ~\ref{app:time:dependent:Hamiltonian}, which is based on the work in~\cite{romero2011optically}. There are several practical ways in which one may envisage to increase the non-linearity by modulating the coupling, depending on the nature of the trap at hand: 
\begin{itemize}
	\item[i)] \textbf{Optically-trapped levitating particles.} The effect that we are looking for can be realised by modulating the phase of the trapping laser beam (which, in turn, can be achieved through an acousto--optical modulator). In our derivation in ~\ref{app:time:dependent:Hamiltonian}, this phase is denoted by $\varphi(\tau)$ and it affects the light--matter coupling strength by determining the particle's location with respect to the standing wave of the cavity field. Thus if we let $\varphi(\tau) = \frac{\pi}{2}\left( 1 + \epsilon \sin{\Omega_0 \tau} \right)$, with $\Omega_0 = \omega_0 / \omega_{\mathrm{m}}$, and where $\omega_0$ is the phase modulation frequency, we obtain the expression used in Section~\ref{sec:time:dependent:coupling}. If, then, the phase frequency is resonant with $\Omega_0 =1$, it should be possible to increase the non-Gaussianity even further. 

	\item[ii)] \textbf{Paul traps.} The shuttling of ions has been demonstrated~\cite{walther2012controlling, hensinger2006t} using Paul traps, which are customarily used for ions but which have also recently been used for trapping nanoparticles~\cite{millen2015cavity, fonseca2016nonlinear, aranas2016split}. These works indicate that a modulation of the particle's position, and hence, a modulation of the coupling as per point i), can be obtained in a Paul trap as well. 

	\item[iii)] \textbf{Micromotion in hybrid traps.} Paul traps display three different kinds of particle motion. Firstly, we have \textit{thermal motion}, whereby the particle moves around the trap. Secondly, and most importantly to our scheme, we have \textit{micromotion}, which induces small movements around the potential minimum. Finally, there is \textit{mechanical motion}, which is the harmonic motion in the trap, here denoted by $\omega_{\mathrm{m}}$. Since the micromotion moves the bead around the potential minimum with a frequency $\omega_{\mathrm{d}}$, this already modulates the light--matter coupling, and is, in a way, an equivalent implementation to the ``shaking'' of the trap. If the micromotion can be engineered to occur with a frequency $\omega_{\mathrm{d}}$ equal to $\omega_{\mathrm{m}}$, then one could, instead of averaging it out, adopt the micromotion's variables to increase the non--Gaussianity with the scheme we propose in Section~\ref{resonance}. To date, the micromotion is generally smaller than the mechanical frequency, $\omega_{\mathrm{d}} \leq \omega_{\mathrm{m}}$, but current experimental efforts appear promising. 
\end{itemize}

There are potentially many more ways in which the light--matter coupling could be modulated, including with optomechanically induced transparecny~\cite{weis2010optomechanically, karuza2013optomechanically} and by using the Kerr effect to change the refractive index of the oscillator. 

We conclude that the enhancement of the non-linearity predicted by our work can be realised in experiments, given the capabilities mentioned above. There are, of course, many challenges to be overcome. In fact, to take advantage of the rather slow logarithmic scaling with time $\tau$, one must keep the system coherent for longer, which is difficult. However, although our analytical results are restricted to Hamiltonian systems, we note that there is no reason to expect that this enhancement should disappear in a noisy setting.

\subsection{Detecting and measuring non-Gaussianity in optomechanical systems}\label{subsec:detecting}
In practise, how would one proceed to measure the amount of non-Gaussianity in the laboratory? As shown in~\cite{barbieri2010non}, the measure of non-Gaussianity used in this work has been measured for the addition of single photons to a coherent state. This requires full state tomography and is thus an expensive process. There are however others ways to proceed. In~\cite{hughes2014quantum} a witness of non-Gaussianity was proposed based on bounding the average photon number in the system from above. While they apply to a single system, they can probably be extended to bipartite systems as well. 

Finally, we here suggest a simple method by which non-Gaussianity can be detected for pure states. We note that the von Neumann entropy $S(\hat{\rho}_{AB})$ of a bipartite state $\hat{\rho}_{AB}$ is bounded by $S(\hat{\rho}_{AB})\geq |S(\hat{\rho}_{A})-S(\hat{\rho}_{B})|$, through the Araki--Lieb inequality~\cite{araki2002entropy} where $\hat{\rho}_{A}$ and $\hat{\rho}_{B}$ are the reduced states of the optical and mechanical subsystems, respectively. Therefore, the measure of non-Gaussianity $\delta(\tau)$ that we defined in~\eqref{measure:of:non:gaussianity} is lower-bounded by
\begin{align}\label{bound:on:measure:of:non:gaussianity}
\delta(\tau)\geq|S(\hat{\rho}_{A})-S(\hat{\rho}_{B})|-S(\hat{\rho}(0)) . 
\end{align}
In this sense, this reduced measure acts as a sufficient (but not necessary) condition for non-Gaussianity. That is, finding that the measure is non-zero does tell us that the state is non-Gaussian, however it does not tell us the full magnitude of the non-Gaussianity. Furthermore, to compute this measure, one would still have to measure the second moments of the optical and mechanical subsystems. This does, however, require fewer measurements than full state tomography on the joint optical and mechanical system. 

\section{Conclusions}\label{sec:conclusions}

We have quantified the non-Gaussianity of initially Gaussian coherent states evolving under the standard, time-dependent optomechanical Hamiltonian. We used a measure of non-Gaussianity based on the relative entropy of a state to characterise the deviation from Gaussianity of the full system. Our techniques allowed us to derive asymptotic expressions for small and large optical coherent-state amplitudes, see Equation~\eqref{eq:delta:small} and Equation~\eqref{eq:delta:large} respectively. We found that for coherent states with amplitude $\lvert\mu_{\mathrm{c}}\rvert\geq 1$, the amount of non-Gaussianity grows logarithmically with the input average number of excitations $|\mu_{\mathrm{c}}|$ and with the light--matter coupling. At resonance, we find that the non-Gaussianity is further enhanced by a logarithmic scaling with the time of interaction. 

An important and promising aspect of our study consists in showing that the amount of non-Gaussianity in the system can be continuously increased by driving a time-modulated optomechanical coupling at mechanical resonance. This allows us to circumvent the usual periodic increase and decrease of non-Gaussianity, and we find that this behaviour effectively yields a non-Gaussian steady-state in the presence of noise. As such, this points to a practically accessible, mechanism to enhance the nonlinear character of optomechanical dynamics at a given light--matter coupling strength. We point out that certain systems, such as hybrid-trap systems, are particularly well-suited for this purpose, as their light--matter interaction is naturally modulated due to the trap characteristics. Finally, we also conclude that the mechanical system does not have to be cooled to the ground state in order to access significant amounts of the non-Gaussianity. 

Our work can be extended to more complicated Hamiltonians of bosonic modes, and we can include modifications such as squeezing of the mechanical state. This setting will be explored in future work.

\section*{Acknowledgments}
We thank Fabienne Schneiter, Daniel Braun, Nathana\"{e}l Bullier, Antonio Pontin, Peter F. Barker, Ryuji Takagi, Francesco Albarelli, Marco G. Genoni, James Bateman and the reviewers for helpful comments and discussions. 

SQ acknowledges support from the EPSRC Centre for Doctoral Training in Delivering Quantum Technologies and thanks the University of Vienna for its hospitality. DR would like to thank the Humboldt Foundation for
supporting his work with their Feodor Lynen Research
Fellowship. This work was supported by the European Union's Horizon 2020 research and innovation programme under grant agreement No.\ 732894 (FET Proactive HOT).

\section*{Data availability statement}
The figures in this work were generated using \textit{Mathematica}. The numerical results in Section~\ref{subsec:open:system:constant:coupling} and~\ref{subsec:open:system:resonance} were obtained using the \textit{Python} library \textit{QuTiP}. The code used to generate the data can be found at \href{https://github.com/sqvarfort/QM-Nonlinearities}{https://github.com/sqvarfort/QM-Nonlinearities}.

\section*{References}

\bibliographystyle{iopart-num}
\bibliography{bibliography}

\newpage

\appendix

\section{Derivation of the dynamics and general tools}\label{appendix:one}
In this appendix, we will derive the coefficients in~\eqref{sub:algebra:decoupling:solution:text} that determine the time-evolution of the system. This follows the derivation in~\cite{Bruschi:Xuereb:2018}. We will also show the explicit time-dependence of the second moments and discuss some methods related to computing the symplectic eigenvalues of the covariance matrix. 

\subsection{Properties of of the nonlinear Hamiltonian}\label{appendix:one:one}
Firstly, we remind the reader that the laboratory time $t$ is rescaled by $\omega_{\mathrm{m}}$. Finding a simple expression for $\hat{U}(\tau)$ is straight-forward when the light--matter coupling $\tilde{g} = g/\omega_{\mathrm{m}}$ is not time-dependent. If $\tilde{g}(\tau) = g(\omega_{\mathrm{m}} \tau)/\omega_{\mathrm{m}}$ is time-dependent we require a more rigorous framework. This is what we present here. 

We will here follow the derivation in Appendix A in ~\cite{Bruschi:Xuereb:2018}. We note that compared with~\cite{Bruschi:Xuereb:2018}, we have here swapped the definition of $\hat{a}$ and $\hat{b}$, and we have a minus-sign in front of $\tilde{g}(\tau)$.  

For the time-dependent Hamiltonian $\hat{H}$ in~\eqref{main:time:independent:Hamiltonian:to:decouple}, the time-evolution operator is given by 
\begin{equation}
\label{appendix:time:evolution:operator}
\hat{U}(\tau):=\overset{\leftarrow}{\mathcal{T}}\,\exp\biggl[-i \, \int_0^{\tau} \mathrm{d}\tau '\,\hat{H}(\tau')\biggr],
\end{equation}
where $\overset{\leftarrow}{\mathcal{T}}$ is the time-ordering operator. 

The basis for decoupling the operator is finding a Lie algebra of generators $\hat{G}_i$ that induce the time-evolution. This Lie algebra must be closed under commutation, that is, either $[\hat{G}_j, \hat{G}_k] \propto \hat{G}_l$, or $[\hat{G}_j, \hat{G}_k] = c$ where $c$ is a scalar. This will allow for the terms in $\hat{U}(t)$ to be moved with the Baker--Campbell--Hausdorff formula such that $\hat{U}(t)$ can be written in a simpler form. 

We start with the ansatz that the evolution operator~\eqref{time:evolution:operator} can be written as 
\begin{equation}
\hat{U}(\tau) = \prod_j \hat{U}_j (\tau) = \prod_j e^{- i F_j \, \hat{G}_j}, 
\end{equation}
where $F_j$ are coefficients corresponding to each of the generators $\hat{G}_j$. Our task is now to find the coefficients $F_j$. 

We begin by defining the operators $\hat{G}_j$ in the algebra:
\begin{align}\label{app:basis:operator:Lie:algebra}
	\hat{N}_a &:= \hat a^\dagger \hat a &
	\hat{N}_b &:= \hat b^\dagger \hat b 
	 & \hat{N}^2_a &:= (\hat a^\dagger \hat a)^2\nonumber\\
	\hat{N}_a\,\hat{B}_+ &:= \hat{N}_a\,(\hat b^{\dagger}+\hat b) &
	\hat{N}_a\,\hat{B}_- &:= \hat{N}_a\,i\,(\hat b^{\dagger}-\hat b), &
	 & 
\end{align}
It can be verified that the operators in~\eqref{app:basis:operator:Lie:algebra} form a closed Lie algebra under commutation. With these operators, our ansatz can be written as
\begin{align}\label{appendix:explicit:time:evolution:operator}
\hat U(\tau)= \hat{U}_a(\tau) \, \hat{U}_b(\tau) \, \hat{U}_a^{(2)} (\tau)\, \hat{U}_+(\tau) \, \hat{U}_-(\tau), 
\end{align} 
where we identify
\begin{align}
\hat{U}_a(\tau) &= e^{-i\, F_{\hat N_a} \,  \hat{N}_a} 
&\hat{U}_b(\tau) &= e^{-i\,F_{\hat N_b} \, \hat{N}_b\,}
 &\hat{U}_a^{(2)}(\tau)&= e^{-i\,F_{\hat{N}^2_a}\,\hat{N}^2_a} \nonumber \\
\hat{U}_+(\tau) &= e^{-i\,F_{\hat{N}_a\,\hat{B}_+}\,\hat{N}_a\,\hat{B}_+} 
&\hat{U}_-(\tau) &= e^{-i\,F_{\hat{N}_a\,\hat{B}_-}\,\hat{N}_a\,\hat{B}_-}.
\end{align}
To find the coefficients, we note the following equivalence: 
\begin{align}\label{appendix:time:evolution:operator:equivalence}
\overset{\leftarrow}{\mathcal{T}}\,&\exp\biggl[-i \, \int_0^{\tau} \mathrm{d} \tau'\,\hat{H}(\tau')\biggr] \nonumber \\
&= e^{-i \, F_{\hat{N}_a} \, \hat{N}_a\,}\,e^{-i\,F_{\hat{N}_b} \, \hat{N}_b}\,e^{-i\,F_{\hat{N}^2_a}\,\hat{N}^2_a}\,e^{-i\,F_{\hat{N}_a\,\hat{B}_+}\,\hat{N}_a\,\hat{B}_+} e^{-i\,F_{\hat{N}_a\,\hat{B}_-}\,\hat{N}_a\,\hat{B}_-}.
\end{align}
Differentiating both sides brings down the Hamiltonian~\eqref{main:time:independent:Hamiltonian:to:decouple} on the left, which we here write in terms of the generators~\eqref{basis:operator:Lie:algebra}. We then multiply both sides by $U^{\dag}$ to obtain the following differential equation:
\begin{align} \label{app:equivalence}
\Omega\,&\hat N_a + \,\hat N_b - \tilde{g}(\tau)\,\hat N_a \, \hat{B}_+ \nonumber \\
&= \dot{F}_{\hat{N}_a} \, \hat{N}_a + \dot{F}_{\hat{N}_b} \, \hat{N}_b + F_{\hat{N}_a^2} \, \hat{N}_a^2 + \dot{F}_{\hat{N}_a \, \hat{B}_+} \hat{U}_b (\tau) \, \hat{N}_a \, \hat{B}_+ \hat{U}_b^\dag(\tau) \nonumber \\
&\quad+ \dot{F}_{\hat{N}_a \, \hat{B}_-} \hat{U}_b(\tau) \, \hat{U}_+(\tau) \, \hat{N}_a \, \hat{B}_- \, \hat{U}_+^\dag (\tau) \, \hat{U}^\dag_b(\tau)
\end{align}
where $\dot{F}_i = \frac{d}{dt} F_i$. This is the equation that determines the coefficients. We can now commute all the operators through, where we find
\begin{align}
\hat{U}_b \, \hat{N}_a \, \hat{B}_+ \hat{U}_b^\dag &= \cos{(F_{\hat{N}_b} )} \, \hat{N}_a \, \hat{B}_+ - \sin{(F_{\hat{N}_b})} \hat{N}_a \,  \hat{B}_- \nonumber \\
\hat{U}_b \, \hat{N}_a \, \hat{B}_- \, \hat{U}_b^\dag &= \cos{(F_{\hat{N}_b})} \hat{N}_a \, \hat{B}_- + \sin{(F_{\hat{N}_b})} \, \hat{N}_a \hat{B}_+ \nonumber \\
\hat{U}_+ \, \hat{N}_a \, \hat{B}_- \, \hat{U}_+ ^\dag &= \hat{N}_a \, \hat{B}_- + 2 F_{\hat{N}_a \, \hat{B}_+} \, \hat{N}_a^2, 
\end{align}
By inserting this into~\eqref{app:equivalence}, we are able to determine the coefficients by linear independence. Integrating, we obtain:
\begin{align}\label{appendix:sub:algebra:decoupling:solution:text}
F_{\hat{N}_a }& = \, \Omega\, \tau,  \nonumber \\
F_{\hat{N}_b} &= \, \tau,  \nonumber \\
F_{\hat{N}^2_a}&= 2  \,\int_0^\tau\,\mathrm{d}\tau'\,\tilde{g}(\tau')\,\sin(\tau')\int_0^{\tau'}\mathrm{d}\tau''\,\tilde{g}(\tau'')\,\cos(\tau''),\nonumber\\
F_{\hat{N}_a\,\hat{B}_+}&=- \int_0^\tau\,\mathrm{d}\tau'\,\tilde{g}(\tau')\,\cos(\tau'),\ \text{and}\nonumber\\
F_{\hat{N}_a\,\hat{B}_-}&= -\int_0^\tau\,\mathrm{d}\tau'\,\tilde{g}(\tau')\,\sin(\tau'), 
\end{align}
where $\Omega = \omega_{\mathrm{c}}/\omega_{\mathrm{m}}$ and $\tau = \omega_{\mathrm{m}} \, t$. 
Depending on the form of $\tilde{g}(\tau)$, we can now use these equations to find a simplified form of $\hat{U}(t)$.

\subsection{Computing determinants of symplectic matrices}\label{appendix:one:two}
When computing the amount of non-Gaussianity in ~\eqref{measure:of:non:gaussianity}, it is useful to consider the symplectic eigenvalues of a Gaussian state~\cite{serafini2017quantum}. In short, for an arbitrary covariance matrix $\mathbf{\sigma}$, they are the eigenvalues of the matrix $i \boldsymbol{\Omega} \mathbf{\sigma}$, where $\boldsymbol{\Omega}=\textrm{diag}(-i,-i,i,i)$ is the symplectic form. There are other ways to define the symplectic eigenvalues though. In the following, we have to switch the basis of the operators to a more convenient one, but this does not affect the final result. The correct definition can be found in~\cite{Adesso:Ragy:2014}. Let us write an arbitrary covariance matrix $\boldsymbol{\sigma}$ in the particular basis $\hat{\mathbb{X}}=(\hat{q}_a,\hat{p}_a,\hat{q}_b,\hat{p}_b)^T$ as
\begin{align}
\boldsymbol{\sigma}
=
\begin{pmatrix}
\boldsymbol{A} & \boldsymbol{C} \\
\boldsymbol{C}^T & \boldsymbol{B}
\end{pmatrix},
\end{align}
where $\boldsymbol{A}=\boldsymbol{A}^T$, $\boldsymbol{B}=\boldsymbol{B}^T$ and all matrices are $2\times2$ matrices. The symplectic invariants are defined as the following four quantities: $a^2:=\det(\boldsymbol{A})$, $b^2:=\det(\boldsymbol{B})$, $c_+c_-:=\det(\boldsymbol{C})$, and $\mu^{-2}:=\det(\boldsymbol{\sigma})=(ab-c_+^2)(ab-c_-^2)$.

Finally, we introduce the parameter $\Delta$ as $\Delta:=\det(\boldsymbol{A})+\det(\boldsymbol{B})+2\,\det(\boldsymbol{C})$. The symplectic eigenvalues $\nu_\pm$ are then given by
\begin{align}
2\,\nu_\pm^2:=&\Delta\pm\sqrt{\Delta^2-4\,\textrm{det}(\boldsymbol{\sigma})}\nonumber\\
=&a^2+b^2+2c_+c_-\pm\sqrt{(a^2-b^2)^2+4(a\,c_++b\,c_-)(a\,c_-+b\,c_+)}.
\end{align}

\section{Evolution of first and second moments}\label{appendix:one:time:evolution}
In the Heisenberg picture, the time evolution of the mode operators $\hat{a}$ and $\hat{b}$  induced by the Hamiltonian considered here is simply $\hat a(t):=\hat U^\dag(t)\, \hat a\, \hat U(t)$ and $\hat b(t):=\hat U^\dag(t)\, \hat b\, \hat U(t)$. In terms of the generators of the Lie algebra defined in~\eqref{app:basis:operator:Lie:algebra}, we explicitly have
\begin{align}
\hat a(t):=& e^{-i\,F_{\hat{N}^2_a}}\,e^{-2\,i\,(F_{\hat{N}^2_a}+F_{\hat{N}_a \, \hat{B}_+} \, F_{\hat{N}_a\,\hat{B}_-})\,\hat{N}_a}\,e^{-i\,F_{\hat{N}_a\,\hat{B}_+}\,\hat{B}_+}\,e^{-i\,F_{\hat{N}_a\,\hat{B}_-}\,\hat{B}_-}\hat a\nonumber\\
\hat b(t):=&e^{-i\,\tau}\,\left[\hat b+(F_{\hat{N}_a\,\hat{B}_-}-i\,F_{\hat{N}_a\,\hat{B}_+})\,\hat{N}_a)\right].
\end{align}
This expression can also be rewritten in more compact notation as
\begin{align}
\hat{a}(t) &= e^{- i \theta_a (\hat{N}_a + 1/2)} \,\hat{D}_{\hat{b}}(F^*) \,  \hat{a} \nonumber \\
\hat{b}(t) &= e^{-i\,\tau}\,\left[\hat b+F^* \,\hat{N}_a\right], 
\end{align}
where $\hat D_{\hat b } (F^*)$ is a Weyl displacement operator and where we have introduces the quantities 
\begin{align}
\theta_a &= 2 \, \left( F_{\hat{N}_a^2} + F_{\hat{N}_a \, \hat{B}_+} F_{\hat{N}_a \hat{B}_-} \right) \nonumber \\
F &= F_{\hat{N}_a \, \hat{B}_-} + i F_{\hat{N}_a \, \hat{B}_+}. 
\end{align}
These expectation values can then be used to compute the elements of the covariance matrix $\boldsymbol{\sigma}$, which in our basis are given by 
\begin{align}
\sigma_{11} &= \sigma_{33} = 1 + 2\braket{\hat a^\dag \hat a} - 2 \braket{\hat a^\dag } \braket{\hat a } \nonumber \\
\sigma_{31} &=2 \braket{\hat{a}^2} - 2 \braket{\hat{a}}^2 \nonumber \\
\sigma_{22} &= \sigma_{44} = 1 + 2 \braket{\hat{b}^\dag \hat{b}} - 2\braket{\hat{b}^\dag} \braket{\hat{b}}  \nonumber \\
\sigma_{42}  &= 2 \braket{\hat{b}^2} - 2 \braket{\hat{b}}^2  \nonumber \\
\sigma_{21} &= \sigma_{34} = 2 \braket{\hat{a} \hat{b}^\dag} - 2 \braket{\hat{a}} \braket{\hat{b}^\dag}   \nonumber \\
\sigma_{41} &= \sigma_{32} = 2 \braket{\hat{a} \hat{b}} - 2 \braket{\hat{a} }\braket{\hat{b}},  
\end{align}
where we have suppressed the time-dependence for notational convenience.

We now compute the expectation values for initial optical coherent states and coherent and thermal coherent states of the mechanics. 
\subsection{Mechanical coherent states}\label{appendix:subsubsec;coherent}
For the initial coherent state $\ket{\Psi(t = 0)} = \ket{\mu_{\mathrm{c}}} \otimes \ket{\mu_{\mathrm{m}}}$ in~\eqref{initial:state:two} and ignoring the global phases $e^{- i \, \Omega \,\tau}$, which can be done by transforming into a frame rotating with $\Omega \hat{a}^\dag \hat{a}$, we obtain
\begin{align} \label{app:expectation:values:coherent}
\braket{\hat a (t) } &:= e^{- i \frac{1}{2} \theta_a } \, e^{|\mu_{\mathrm{c}}|^2 (e^{- i \theta_a}-1)} \, e^{-\frac{1}{2}|F|^2} \, e^{F^* \mu_{\mathrm{m}}^* - F \mu_{\mathrm{m}}} \, \mu_{\mathrm{c}} \nonumber \\
\braket{\hat b(t) } &:= e^{- i \tau} \mu_{\mathrm{m}}  +  e^{- i \tau} \, F^* \, |\mu_{\mathrm{c}}|^2  \nonumber \\
\braket{\hat{a}^2(t)} &:= e^{- 2i \, \theta_a} \,   e^{|\mu_{\mathrm{c}}|^2 (e^{- 2 i \theta_a} - 1)} \,  e^{ - 2 |F|^2    } \, e^{2 ( F^* \mu_{\mathrm{m}}^* - F \mu_{\mathrm{m}})} \,  \mu_{\mathrm{c}}^2\nonumber \\
 \braket{\hat b^{2}(t)} &:= e^{-2 i \tau} \left(  \mu_{\mathrm{m}} + F^*\, |\mu_{\mathrm{c}}|^2 \right)^2  + e^{- 2i \tau } F^{*2} \, |\mu_{\mathrm{c}}|^2 \nonumber \\
 \braket{\hat a^\dag(t) \hat a(t) } &:= |\mu_{\mathrm{c}}|^2 \nonumber \\
 \braket{\hat b ^\dag(t) \hat b(t)}  &:=  \left|  \, \mu_{\mathrm{m}}  + F^*\, |\mu_{\mathrm{c}}|^2 \right|^2 + |F|^2 \, |\mu_{\mathrm{c}}|^2\nonumber \\
\braket{\hat a(t) \hat b(t) } &:= e^{- i \frac{1}{2}\theta_a}  \,e^{ |\mu_{\mathrm{c}}|^2 ( e^{- i \theta_a}  - 1)}\, e^{ -\frac{1}{2}  |F|^2   } \, e^{F^* \mu_{\mathrm{m}}^* - F \mu_{\mathrm{m}}} \, \,  \mu_{\mathrm{c}} \, e^{- i \tau}\,  \left[\mu_{\mathrm{m}}  + \left( |\mu_{\mathrm{c}}|^2 e^{- i \theta_a} + 1\right)
F^* \right]  \nonumber \\
\braket{\hat a (t) \, \hat b^\dag(t) }: &=e^{- \frac{1}{2}i \theta_a}  \, e^{ |\mu_{\mathrm{c}}|^2 ( e^{- i \theta_a}  - 1)}  \,e^{- \frac{1}{2}   |F|^2 }\, e^{F^* \mu_{\mathrm{m}}^* - F \mu_{\mathrm{m}}} \,  \mu_{\mathrm{c}} \,  e^{i \tau} \,  \left[  \mu_{\mathrm{m}}^* + |\mu_{\mathrm{c}}|^2 e^{- i \theta_a}  F \right]. 
\end{align}
where we have introduced $F := F_{\hat N_a \, \hat B_- } + i F_{\hat N_a \, \hat B_+}$ and $\theta_a := 2 ( F_{\hat N_a^2} + F_{\hat N_a \, \hat{B}_+} F_{\hat N_a \, \hat B_-} )$. The covariance matrix elements are given by 
\begin{align}\label{full:elements:covaraince:matrix}
\sigma_{11} = \sigma_{33} &= 1 + 2|\mu_{\mathrm{c}}|^2 \left( 1 - e^{-4 \, |\mu_{\mathrm{c}}|^2 \sin^2{\theta_a/2}} \, e^{- |F|^2} \, e^{F^* \mu_{\mathrm{m}}^* - F \mu_{\mathrm{m}}} \right) \nonumber \\
\sigma_{31} &=  2 \, \mu_{\mathrm{c}}^2 \, e^{- i \theta_a} \, e^{- |F|^2} \, \biggl(  e^{- i \theta_a} \, e^{|\mu_{\mathrm{c}}|^2  \, ( e^{-  2 \, i \theta_a} - 1)} \, e^{- \,|F|^2} \, e^{2(F^* \mu_{\mathrm{m}}^* - F \mu_{\mathrm{m}})} \nonumber \\
&\quad\quad- e^{2|\mu_{\mathrm{c}}|^2 ( e^{- i \theta_a} - 1)} \,  e^{2(F^* \mu_{\mathrm{m}}^* - F \mu_{\mathrm{m}})}\biggr) \nonumber \\
\sigma_{22} =  \sigma_{44} &= 1  + 2 \, |\mu_{\mathrm{c}}|^2 \, |F|^2  \nonumber \\
\sigma_{42}  &= 2 \, e^{- 2 \, i \,\tau} \,  |\mu_{\mathrm{c}}|^2 \, F^{*2}  \nonumber \\
\sigma_{21} = \sigma_{34} &=  2  \, F\,\mu_{\mathrm{c}}\, |\mu_{\mathrm{c}}|^2 \, (e^{- i \theta_a} - 1) \, e^{- i \frac{1}{2} \theta_a} \,  e^{i\,\tau} \, e^{|\mu_{\mathrm{c}}|^2 ( e^{- i \theta_a} - 1)} e^{- \frac{1}{2}|F|^2} \, e^{F^* \mu_{\mathrm{m}}^* - F \mu_{\mathrm{m}}} \,  \nonumber \\
\sigma_{41}= \sigma_{32}  &= 2\,  F^*\,\mu_{\mathrm{c}} \,  \left(|\mu_{\mathrm{c}}|^2 (e^{- i \theta_a} -1) + 1  \right) \,  e^{- \frac{i}{2} \theta_a}  \, e^{- i\, \tau} \,e^{ |\mu_{\mathrm{c}}|^2 ( e^{- i \theta_a}  - 1)} \, e^{- \frac{|F|^2}{2} }\,e^{F^* \mu_{\mathrm{m}} ^* - F \mu_{\mathrm{m}}}.
\end{align}

\subsection{Mechanical thermal coherent states}\label{appendix:subsubsec;thermal:coherent}

In Section~\ref{sec:initial:state} we noted that the mechanical state is most often found in a thermal state. We assume that the initial state is a coherent thermal state of the form 
\begin{equation}
\hat \rho_{\mathrm{th}} = \frac{1}{\bar{n} \pi} \int \mathrm{d}^2 \beta \, e^{- |\beta|^2/\bar{n}} \ket{\beta}\bra{\beta}, 
\end{equation}
where $\bar{n}$ is the average thermal phonon occupation number. The cavity is still in the coherent state $\ket{\mu_{\mathrm{c}}}$. Several of the expectation values can then simplified by noting that
\begin{align}
&\int \mathrm{d}^2 \beta \, \beta = 0,  &&\int \mathrm{d}^2 \beta \, \beta^2 = 0.
\end{align}
As a result, the expectation values for the initial mechanical states as a coherent thermal state are given by 
\begin{align}\label{app:expectation:values:coherent:thermal}
\braket{\hat a (t) } &:= \frac{1}{\bar{n}\pi} \int \mathrm{d}^2 \beta \, e^{- |\beta|^2/\bar{n}} \,  e^{- i \frac{1}{2} \theta_a } \, e^{|\mu_{\mathrm{c}}|^2 (e^{- i \theta_a}-1)} \, e^{-\frac{1}{2}|F|^2} \, e^{F^* \beta^* - F \beta} \, \mu_{\mathrm{c}} \nonumber \\
\braket{\hat b(t) } &:=   e^{- i \tau} \, F^* \, |\mu_{\mathrm{c}}|^2\nonumber \\
\braket{\hat{a}^2(t)} &:= \frac{1}{\bar{n} \pi} \int \mathrm{d}^2 \beta \, e^{- |\beta|^2/\bar{n}} \,  e^{- 2i \, \theta_a} \,   e^{|\mu_{\mathrm{c}}|^2 (e^{- 2 i \theta_a} - 1)} \,  e^{ - 2 |F|^2    } \, e^{2 ( F^* \beta^* - F \beta)} \,  \mu_{\mathrm{c}}^2\nonumber \\
 \braket{\hat b^{2}(t)} &:=  e^{- 2 \, i \, \tau} F^{*2} \, |\mu_{\mathrm{c}}|^2 \left( 1 +
 |\mu_{\mathrm{c}}|^2 \right)   \nonumber \\
 \braket{\hat a^\dag(t) \hat a(t) } &:= |\mu_{\mathrm{c}}|^2 \nonumber \\
 \braket{\hat b ^\dag(t) \hat b(t)}  &:= |F|^2 |\mu_{\mathrm{c}}|^2 \left( 1 +  |\mu_{\mathrm{c}}|^2 \right) \nonumber \\
\braket{\hat a(t) \hat b(t) } &:= \frac{1}{\bar{n} \pi} \int \mathrm{d}^2 \beta \, e^{- |\beta|^2/\bar{n}} \, e^{- i \frac{1}{2}\theta_a}  \,e^{ |\mu_{\mathrm{c}}|^2 ( e^{- i \theta_a}  - 1)}\, e^{ -\frac{1}{2}  |F|^2   } \, e^{F^* \beta^* - F \beta} \, \,  \mu_{\mathrm{c}} \nonumber \\
&\quad\quad\quad\quad  \times  \, e^{- i \, \tau}\,  \left[\beta  + \left( |\mu_{\mathrm{c}}|^2 e^{- i \theta_a} + 1\right)
F^* \right]  \nonumber \\
\braket{\hat a (t) \, \hat b^\dag(t) }: &=\frac{1}{\bar{n}\pi} \int \mathrm{d}^2 \beta \, e^{- |\beta|^2/\bar{n}} \, e^{- \frac{1}{2}i \theta_a}  \, e^{ |\mu_{\mathrm{c}}|^2 ( e^{- i \theta_a}  - 1)}  \,e^{- \frac{1}{2}   |F|^2 }\, e^{F^* \beta^* - F \beta} \,  \mu_{\mathrm{c}} \, \nonumber \\
&\quad \quad \quad \quad \times e^{i \, \tau} \,  \left[  \beta^* + |\mu_{\mathrm{c}}|^2 e^{- i \theta_a}  F \right]. 
\end{align}
The resulting covariance matrix elements can be computed from here.

\section{Derivation of the asymptotic form of the symplectic eigenvalues}\label{section:symplectic:eigenvalues}
The symplectic eigenvalues $\nu_\pm$ have the expression $\nu_\pm=1+\delta\nu_\pm$. We would like to see what is the form of the function $f(x)=\frac{x+1}{2}\ln\frac{x+1}{2}-\frac{x-1}{2}\ln\frac{x-1}{2}$ when we compute $f(\nu_\pm)$ and $\delta\nu_\pm\ll1$ or $\delta\nu_\pm\gg1$. 

In the first case, $\delta\nu_\pm\ll1$, and we have
\begin{align}\label{small:temperature:expansion:appendix}
f(\nu_+)=&f(1+\delta\nu_+)\nonumber\\
=&\frac{2+\delta\nu_+}{2}\ln\frac{2+\delta\nu_+}{2}-\frac{\delta\nu_+}{2}\ln\frac{\delta\nu_+}{2}\nonumber\\
=&\left(1+\frac{\delta\nu_+}{2}\right)\ln\left(1+\frac{\delta\nu_+}{2}\right)-\frac{\delta\nu_+}{2}\ln\frac{\delta\nu_+}{2}\nonumber\\
=&\left(1+\frac{\delta\nu_+}{2}\right)\frac{\delta\nu_+}{2}-\frac{\delta\nu_+}{2}\ln\frac{\delta\nu_+}{2}+\mathcal{O}\left(\left(\frac{\delta\nu_+}{2}\right)^3\right)\nonumber\\
=&-\frac{\delta\nu_+}{2}\ln\frac{\delta\nu_+}{2}+\mathcal{O}\left(\frac{\delta\nu_+}{2}\right).
\end{align}
An analogous computation can be done for $\nu_-$. The last line of~\eqref{small:temperature:expansion:appendix} is a consequence of the fact that $-x\ln x\gg x$ for $x\ll1$.

In the second case we have $\delta\nu_\pm\gg1$, therefore
\begin{align}\label{large:temperature:expansion:appendix}
f(\nu_+)=&f(1+\delta\nu_+)\nonumber\\
=&\frac{2+\delta\nu_+}{2}\ln\frac{2+\delta\nu_+}{2}-\frac{\delta\nu_+}{2}\ln\frac{\delta\nu_+}{2}\nonumber\\
=&\left(1+\frac{\delta\nu_+}{2}\right)\ln\left(1+\frac{\delta\nu_+}{2}\right)-\frac{\delta\nu_+}{2}\ln\frac{\delta\nu_+}{2}\nonumber\\
=&\left(1+\frac{\delta\nu_+}{2}\right)\ln\frac{\delta\nu_+}{2}\left(1+\frac{2}{\delta\nu_+}\right)-\frac{\delta\nu_+}{2}\ln\frac{\delta\nu_+}{2}\nonumber\\
=&\left(1+\frac{\delta\nu_+}{2}\right)\ln\frac{\delta\nu_+}{2}+\left(1+\frac{\delta\nu_+}{2}\right)\ln\left(1+\frac{2}{\delta\nu_+}\right)-\frac{\delta\nu_+}{2}\ln\frac{\delta\nu_+}{2}\nonumber\\
=&\ln\frac{\delta\nu_+}{2}+\left(1+\frac{\delta\nu_+}{2}\right)\ln\left(1+\frac{2}{\delta\nu_+}\right)+\mathcal{O}\left(\left(\frac{2}{\delta\nu_+}\right)^2\right)\nonumber\\
=&\ln\frac{\delta\nu_+}{2}+\left(1+\frac{\delta\nu_+}{2}\right)\,\frac{2}{\delta\nu_+}+\mathcal{O}\left(\left(\frac{2}{\delta\nu_+}\right)^2\right)\nonumber\\
=&\ln\frac{\delta\nu_+}{2}+1+\mathcal{O}\left(\frac{2}{\delta\nu_+}\right),
\end{align}
which concludes the proof of the claim, since $\delta\nu_\pm\gg1$ and therefore $\ln\frac{\delta\nu_+}{2}\gg1$. An analogous computation can be done for $\nu_-$.

\section{Coefficients for time-dependent light--matter coupling} \label{app:time:dependent:coefficients}
In this appendix, we compute the coefficients used in Section~\ref{sec:time:dependent:coupling}. Starting from~\eqref{appendix:sub:algebra:decoupling:solution:text}, we assume that the coupling has the functional form $\tilde{g}(\tau) = \tilde{g}_0 ( 1 + \epsilon \sin{ \tau \, \Omega_0})$, where we have set $\tilde{g}(\tau) = g(\omega_{\mathrm{m}}t )/\omega_{\mathrm{m}}$. The algebra is straightforward, although cumbersome, and it leads us to the expressions:
\begin{align} \label{eq:time:dependent:coefficients}
F_{\hat{N}_a^2} &= -\tilde{g}_0^2
\bigl[ \tau -\sin (\tau) \cos (\tau) \bigr] +2 \, \epsilon \frac{\tilde{g}_0^2}{ \Omega_0 }  \biggl[ \sin ^2( \tau ) \cos (\tau  \Omega )- 2 \sin^2 \left(\frac{\tau}{2} \right)\biggr]\nonumber \\
&- \epsilon \frac{\tilde{g}_0^2}{4 \Omega_0 ( 1 + \Omega_0)} \, 4  \sin (2\tau ) \sin ( \Omega_0 \, \tau )- \epsilon \frac{\tilde{g}_0^2}{2 \Omega_0 ( 1 - \Omega_0^2)} \, 8 \cos (\tau ) \sin^2\left( \frac{(1 - \Omega_0)\tau}{2} \right)
\nonumber \\
&+ \epsilon^2 \, \frac{\tilde{g}_0^2}{ 4 \, \Omega_0 ( 1 + \Omega_0)} \, \left( 2 \, \tau-4 \sin (\tau) \cos ( \Omega_0 \, \tau ) (\cos (\tau) \cos ( \Omega_0 \, \tau )-2)\right) \nonumber \\
&+ \epsilon^2 \, \frac{\tilde{g}_0^2}{4 \, \Omega_0 ( 1 - \Omega^2_0)} \, \biggl( 4 \, \sin (\tau) \cos ( \Omega_0 \, \tau ) (\cos (\tau) \, \cos ( \Omega_0 \, \tau )-2)+8 \cos (\tau) \,  \sin ( \Omega_0 \, \tau )\nonumber \\
&\quad\quad\quad\quad\quad\quad\quad\quad\quad\quad+(1-2 \, \cos (2\, \tau)) \sin (2\,  \Omega_0 \, \tau ) - 2 \, \tau\biggr) \nonumber \\
&+ \epsilon^2 \, \frac{\tilde{g}_0^2}{4 \, \Omega_0 \, (1 - \Omega_0^2)^2} \biggl( 8 \, \Omega_0 \,   \sin (\tau)\,  \cos ( \Omega_0 \, \tau )-2 \, \Omega_0\,  \sin (2 \, \tau) \, \cos (2 \, \tau\,  \Omega_0 ) \nonumber \\
&\quad\quad\quad\quad\quad\quad\quad\quad\quad\quad-8 \cos (\tau) \sin (\Omega_0 \, \tau)+2 \cos (2 \, \tau) \, \sin (2 \, \tau\,  \Omega_0 ) \biggr) \nonumber \\
F_{\hat N_a \, \hat{B}_+} &= -\frac{\tilde{g}_0}{1 + \Omega_0} \, \epsilon \sin(\tau) \sin(\Omega_0 \, \tau) + \frac{2 \, \Omega_0 \, \tilde{g}_0}{1 - \Omega_0^2} \, \epsilon \, \sin^2 \left( \frac{( 1 - \Omega_0)\tau}{2} \right) -\tilde{g}_0 \, \sin (\tau)\nonumber \\
F_{\hat N_a \, \hat{B}_- } &= - \frac{\tilde{g}_0 }{1 + \Omega_0} \, \epsilon \, \left(  \sin (\tau) \, \cos ( \Omega_0 \, \tau) + \sin((1 + \Omega_0)\tau) \right)+ \frac{\tilde{g}_0}{1 - \Omega_0^2} \, \epsilon \, \sin((1 - \Omega_0)\tau) \nonumber \\
&\quad \quad- 2 \, \tilde{g}_0 \, \sin^2 \left( \frac{\tau}{2} \right)
\end{align}
It can be seen from these expressions that there are some resonances expected, namely a drastic change in the behaviour of (some of) the functions in the limit $\Omega_0\rightarrow1$, which occurs when $\omega_0 = \omega_{\mathrm{m}}$. 

It is straight-forward to see how the terms $F_{\hat{N}_a \, \hat{B}_+}$ and $F_{\hat{N}_a \, \hat{B}_-}$ simplify as $\Omega_0 \rightarrow 0$ by noting that $\lim_{\Omega_0 \rightarrow 1 } \sin{\tau( 1 - \Omega_0}/(1 - \Omega_0^2) = \tau/2$. The long expression for $F_{\hat{N}_a^2} $ is more challenging. We note that the terms independent of $\epsilon$ remain unchanged with $\Omega_0$. Thus, at resonance, these coefficients read:
\begin{align}
F_{\hat N_a^2} &=-\frac{1}{32} \tilde{g}_0 \, \bigl[ \epsilon ^2 \,   \bigl( 12 \, \tau-8 \sin (2\, \tau)+\sin (4 \, \tau) \bigr)\nonumber \\
&\quad\quad\quad\quad\quad\quad\quad\quad+ \epsilon \, (64-72 \cos (\tau)+8 \cos (3 \, \tau)) + 32 \, \tau-16 \sin (2\, \tau) \bigr]\nonumber\\
F_{\hat N_a \, \hat B_+} &= -\tilde{g}_0 \sin (\tau) \left(1+ \frac{\epsilon}{2} \sin (\tau)\right) \nonumber \\
F_{\hat N_a \, \hat B_-} &=\frac{\tilde{g}_0}{4} \epsilon   \,  \left(  \sin (2 \, \tau)-2 \, \tau  \right) -  2 \, \tilde{g}_0 \,  \sin^2\left( \frac{\tau}{2} \right). 
\end{align}
\section{Derivation of the modulated light--matter coupling} \label{app:time:dependent:Hamiltonian}
In this appendix, we will show how the time-dependent term used in Section~\ref{sec:time:dependent:coupling} can be derived for levitated nanobead systems.  In~\cite{romero2011optically}, a fully general theory of light--matter coupling is presented. We will recount some of the derivation here and show how the cavity volume can be modulated in a manner such that it is useful to our scheme. 

Given a number of assumptions regarding the light--matter interaction (see~\cite{romero2011optically} for a full description) the full Hamiltonian that describes the light--matter interaction for a homogeneous dielectric object is the following: 
\begin{equation}
\hat{H}^{\mathrm{tot}} = \hat{H}^f_m + \hat{H}^f_c + \hat{H}^f_c + \hat{H}^f_{\mathrm{out}} + \hat{H}^f_{\mathrm{free}} + \hat{H}^i_{\mathrm{cav-out}} + \hat{H}^i _{\mathrm{diel}}. 
\end{equation}
The term $\hat{H}^f_m = \hat{p}^2/2M$, where $M$ is the total mass of the system, is the kinetic energy of the centre-of-mass position along the cavity axis. $\hat{H}_c^F = \hbar \omega_{\mathrm{c}}\hat{a}^\dag \hat{a}$ is the energy of the cavity mode. $\hat{H}^f_{\mathrm{out}}$ and $\hat{H}^f_{\mathrm{free}}$ are terms describing an open system, which we shall ignore in this work. We likewise ignore $\hat{H}^i_{\mathrm{cav-out}}$ which describes a coupling between the cavity input and the output mode. 

The last term $\hat{H}^i_{\mathrm{diel}}$ describes the light--matter coupling and can be written in the general form
\begin{equation} \label{app:def:dielectric:hamiltonian}
\hat{H}^i_{\mathrm{diel}} = - \frac{1}{2}\int_{V(\boldsymbol{r})} \mathrm{d} \boldsymbol{x} \, P(\boldsymbol{x}) \, \hat{E}(\boldsymbol{x}), 
\end{equation}
where $P(\boldsymbol{x})$ is the polarization of the levitated objects (which we assume to be a scalar quantity) and $\hat{E}(\boldsymbol{x})$ is the total electric field, which can be obtained from solving Maxwell's equations given a set of well-defined boundary conditions. The quantised modes of the electric field can thus be written as~\cite{serafini2017quantum}
\begin{equation} \label{app:def:electric:field}
\hat{E}(\boldsymbol{x}) = i \sum_{s,m} E_{m} \left( a_{s, m}\,  - a^\dag_{s,m} \right) \chi_{s,m}(\boldsymbol{x}), 
\end{equation}
where $s$ is the spin-polarization index and $m$ signifies the field-mode number, and $E_m = \sqrt{\frac{\omega_{\mathrm{m}} \hbar}{2 \epsilon_0 \, V_{\mathrm{c}}}}$ is the field amplitude with $V_{\mathrm{c}}$ being the cavity mode volume. The functions $\chi_{s,m} $ must obey the spatial solutions to the wave-equations, where the full classical solutions separate into $\boldsymbol{E}( \boldsymbol{r}, t) = \chi ( \boldsymbol{r}) \, T(t)$. 

If we assume that the polarization is given by $P(\boldsymbol{x}) = \epsilon_{\mathrm{c}} \epsilon_0  E(\boldsymbol{x})$, we obtain the simpler expression 
\begin{equation} \label{app:def:dielectric:hamiltonian:simple}
\hat{H}^i _{\mathrm{diel}} = - \frac{\epsilon_{\mathrm{c}} \epsilon_0}{2} \int_{V(\boldsymbol{r})} \, \mathrm{d} \boldsymbol{x}\, [\hat{E}(\boldsymbol{x})]^2, 
\end{equation}
where  $\epsilon_{\mathrm{c}} = 3 \frac{ \epsilon_r - 1}{\epsilon_r + 2}$, and where $\epsilon_r$ is the relative dielectric constant of the nanodiamond. 

We now assume that the electric field operators are displaced by a classical part: $\hat{a} \rightarrow \braket{\hat{a}_0} + \hat{a}$. The classical part $ \braket{\hat{a}_0}$ will form the optical trapping field, while the quantum part describes the light--matter interaction. 

Thus the classical contribution to the electrical field is given by 
\begin{equation}
\mathcal{E}(\boldsymbol{x}) = i \sqrt{\frac{\omega_{\mathrm{c}}}{2\epsilon_0 V_{\mathrm{c}}}} \left( \alpha f(\boldsymbol{x}) - \alpha^* f^*(\boldsymbol{x}) \right), 
\end{equation}
where $\alpha$ is a complex prefactor and $f(\boldsymbol{x})$ is a complex function which describe the standing waves inside the cavity. We now write our full electric field as $\hat{E}_{\mathrm{tot}}(\boldsymbol{x}) = \hat{E}(\boldsymbol{x}) + \mathcal{E}(\boldsymbol{x})$, where $\hat{E}(\boldsymbol{x})$ is the quantum contribution containing $\hat{a}$ and $\hat{a}^\dag$, and $\mathcal{E}(\boldsymbol{x})$ is the classical part.  The full Hamiltonian is now 
\begin{align}
 \hat{H}^i _{\mathrm{diel}} &= - \frac{\epsilon_{\mathrm{c}} \epsilon_0}{2} \int_{V(\boldsymbol{r})} \, \mathrm{d} \boldsymbol{x}\, [\hat{E}(\boldsymbol{x}) + \mathcal{E}(\boldsymbol{x})]^2 \nonumber \\
 &=  - \frac{\epsilon_{\mathrm{c}} \epsilon_0}{2} \int_{V(\boldsymbol{r})} \, \mathrm{d} \boldsymbol{x}\, [\hat{E}^2(\boldsymbol{x}) + \mathcal{E}^2(\boldsymbol{x}) + 2\hat{E}(\boldsymbol{x}) \mathcal{E} (\boldsymbol{x})]. 
\end{align}
The classical contribution, $\mathcal{E}(\boldsymbol{x})$ will yield a trapping frequency, while the operator terms $\hat{E}(\boldsymbol{x})$ will yield the light--matter interaction term for the levitated sphere. The cross-term, $\hat{E}(\boldsymbol{x}) \mathcal{E}^2 (\boldsymbol{x})$ will generate elastic scattering processes inside the cavity which converts cavity photons and tweezer photons into free modes~\cite{romero2011optically}. We shall ignore them here and focus on the generation of the trapping frequency $\omega_{\mathrm{m}}$ and the coupling $g(t)$. We begin with the trapping frequency. 

\subsection{Mechanical trapping frequency}
We now assume that the classical field has a Gaussian profile which extends in the $y$-direction for a cylindrical geometry. The cavity extends along the $z$-direction. We here follow the derivation presented in~\cite{clemente2010magnetically}. 

If we denote the radius of the cylinder by $r$, we can write down the trapping field as 
\begin{equation}
\mathcal{E}(y, r) = E_0 \frac{W_0}{W(y)} \exp{\left( - \frac{r^2}{W^2(y)} \right)}, 
\end{equation}
where $E_0 = \sqrt{\frac{P_t}{\epsilon_0 c \pi W_0^2} }$, $P_t$ is the trapping laser power and $W_0$ is the beam waist with the full beam as a funtion of $y $ being $W(y) = W_0 \sqrt{1 + \frac{y^2 \lambda^2}{\pi^2 W_0^4}}$. It  follows that the narrowest part of the beam $W_0$ occurs at $y = 0$, which is the minimum in the potential where the nanobead is trapped. 

We can now expand $[\mathcal{E}(y,r)]^2$ to second order in $r$ and $y$ around the origin $y_0 = r_0 = 0$. We start with the exponential, which we expand as
\begin{equation}
[\mathcal{E} (y,r)]^2 \approx E_0^2 \frac{W_0^2}{W^2(y)} \left( 1 - 2\frac{r^2}{W^2(y)} \right). 
\end{equation}
Next, we expand the inverse beam width to second order in $y$:
\begin{equation}
\frac{1}{W^2(y)} \approx  \frac{1}{W_0^2} \left( 1 - \frac{y^2 \lambda^2}{2 \pi^2 W_0^4} \right). 
\end{equation}
Combining the two expressions give us
\begin{align}
[\mathcal{E}(y, r)]^2 &\approx E_0^2 \left( 1  - \frac{y^2 \lambda^2}{2\pi^2 W_0^4} \right) \left( 1 - 2 \frac{r^2}{W_0^2} \left( 1 - \frac{y^2 \lambda^2}{2 \pi^2 W_0^2} \right) \right) \nonumber \\
&\approx E_0^2 - \frac{E_0^2 \, y^2 \, \lambda^2 }{\pi^2 \, W_0^2} + r^2 E_0^2 \left( \frac{4 y^2 \, \lambda^2}{\pi^2 \, W_0^4} - \frac{2}{W_0^2}\right). 
\end{align}
If we now assume that $y \ll W_0$, meaning that the beam waist is much larger than the region we consider, we can approximate the above as
\begin{equation}
[\mathcal{E}(y,r)]^2 \approx E_0^2 - r^2E_0^2 \frac{2}{W_0^2}. 
\end{equation}
We then insert this now constant expression into the integral for the Hamiltonian and we drop all constant terms as they are just constant energy shifts. To perform this integral, we now assume that the radius $R$ of the bead is much smaller than the wavelength of the light. This is often referred to as the `point--particle approximation', or the Rayleigh approximation. Essentially, this means that the field inside the bead is constant (although the field still changes in space with $x$ and $y$). Thus we can assume that wherever the sphere is located in the field, the integral just simplifies to the volume of the sphere times the field amplitude. For a derivation which includes arbitrary particle sizes, see~\cite{pflanzer2012master}. 

This gives 
\begin{align}
H_{\mathrm{trap}} \approx   \frac{\epsilon_{\mathrm{c}}}{2} \int_{V(\boldsymbol{r})} \mathrm{d} \boldsymbol{x} \, r^2 E_0^2 \frac{2}{W_0^2} \approx r^2\frac{\epsilon_{\mathrm{c}} E_0^2 }{W_0^2} V, 
\end{align}
where $V$ is the integration volume. The result is a harmonic trapping of the form 
\begin{equation}
\frac{1}{2} m \omega_{\mathrm{m}}^2 r^2 = \frac{\epsilon_{\mathrm{c}} E_0^2 }{W_0^2} Vr^2, 
\end{equation}
where we identify the trapping frequency as 
\begin{equation}
 \omega_{\mathrm{m}}^2 = \frac{2}{m} \frac{\epsilon_{\mathrm{c}} E_0^2 }{W_0^2}V = \frac{12 I m }{\rho c \epsilon_{\mathrm{c}} W_0^2} \left( \frac{\epsilon_r - 1}{\epsilon_r + 2} \right), 
\end{equation}
where $\rho = \frac{m}{V}$ is the density of the levitated object and where we have used $E_0^2 = \frac{2 I }{c \epsilon_0 }$, where $I$ is the intensity of the laser beam, and $\epsilon_{\mathrm{c}} = 3 \frac{ \epsilon_r - 1}{\epsilon_r + 2}$. 

\subsection{The light--matter interaction term}
We now come to the most important term, which is the light--matter interaction term denoted $g$ in this work. We will continue to follow the derivation in~\cite{romero2011optically} to show exactly where time--dependence could potentially be included. 

If the sphere is sufficiently small, we can choose a TEM 00 (transverse electromagnetic mode) as the cavity mode, which is aligned in the $z$-direction. In this mode, the cross-section in $x$ and $y$ is perfectly Gaussian, and it is one of the most commonly used modes in experiments. If the sphere is smaller than the laser waist and if it is placed close to the centre of the cavity, we can approximate the field at the centre of the beam by 
\begin{equation}
[E(\boldsymbol{x})]^2 \approx \frac{\omega_{\mathrm{c}}}{2 \epsilon_0 V_c} \left( 1 - \frac{2( x^2 + y^2)}{W_c^2} \right) \cos^2{(k_c z - \varphi)} \, \hat{a}^\dag \hat{a}. 
\end{equation}
Here, the laser waist is given by $W_c = \sqrt{\frac{\lambda L}{(2 \pi)^2}}$,  $L$ is the cavity length. $\lambda$ is the laser wavelength. We assume that the wave-vector $\boldsymbol{k}_c$ points in the $z$-direction, along the axis of the cavity, and $\varphi$ is a generic phase which determines the minimum of the potential seen by the bead. For laser-trapped nanobeads, this phase can be made time-dependent, whereas for a Paul trap, it is static. We will leave out the time-dependence for now for notational simplicity. Finally, $\hat{a}$ and $\hat{a}^\dag$ are the annihilation and creation operators of the electromagnetic field. 

To obtain the Hamiltonian term, we now integrate over the full energy within the volume of the nanobead. For a bead situated at  $\boldsymbol{r} = (x,y,z)$ leads to 
\begin{align}
\hat{H}_{\mathrm{diel}} &= - \frac{\epsilon_c \epsilon_0}{2} \int_{V(\boldsymbol{r})} \mathrm{d} \boldsymbol{x} \, [E(\boldsymbol{x})]^2 \nonumber \\
&= - \frac{\epsilon_c \epsilon_0}{2} \int_{V(\boldsymbol{r})} \mathrm{d} \boldsymbol{x} \, \frac{\omega_{\mathrm{c}}}{2 \epsilon_0 V_c} \left( 1 - \frac{2( x^2 + y^2)}{W_c^2} \right) \cos^2{(k_c z - \varphi)} \hat{a}^\dag \hat{a}. 
\end{align}
We now assume that the radius of the sphere $R$ is much smaller than the wavelength of the light, such that $k_c R \ll 1$. As mentioned above, this is the `point--particle approximation', or the Rayleigh approximation. 

Thus the integral simplifies to 
\begin{equation}
\hat{H}_{\mathrm{diel}}  =  - \frac{\epsilon_c \epsilon_0}{2} \int_{V(\boldsymbol{r})} \mathrm{d} \boldsymbol{x} \, \frac{\omega_{\mathrm{c}}}{2 \epsilon_0 V_c} \left( 1 - \frac{2( x^2 + y^2)}{W_c^2} \right) \cos^2{(k_c z - \varphi)} \, \hat{a}^\dag \hat{a} = \omega_c f(\boldsymbol{r}) \,  \hat{a}^\dag \hat{a}, 
\end{equation} 
where we have defined the function $f(\boldsymbol{r})$ as 
\begin{align} \label{def:frequency:function}
f(\boldsymbol{r}) = - \frac{V \epsilon_c}{4 V_c } \left( 1 - \frac{2 (x^2 + y^2)}{W_c^2}\right) \cos^2{(k_c z - \varphi)}. 
\end{align}
Now, we assume that the sphere is trapped at position $\boldsymbol{r}_0 = (x_0, y_0, z_0)^{\mathrm{T}}$, which we take to be the origin of the cavity with $x_0 = 0$, $y_0 = 0$ and $z_0 = 0$. For small perturbations to $z$, which we will later quantize, we can expand~\eqref{def:frequency:function} around $z_0 = 0$ to first order. For this to be valid, we must also expand $\varphi$ to first order. We write
\begin{align}
\cos^2{(k_c z - \varphi)} &= [\cos(k_c z) \cos(\varphi) + \sin(k_c z) \sin(\varphi)]^2\nonumber \\
&\approx \left[ \left( 1 - \frac{k_c^2 z^2}{2} \right) \left( 1 - \frac{\varphi^2}{2} \right) + k_c z \varphi \right]^2 \nonumber \\
&\approx \left[ 1 + k_c z \varphi \right]^2 \nonumber \\
&\approx 1 +2 k_c z \varphi . 
\end{align}
We note the linearised $z$-coordinate here, which will later become our quantum operator. We can then write down the full expression
\begin{equation}
\omega_c f(\boldsymbol{r}) \hat{a}^\dag \hat{a} = -\omega_c \frac{V \epsilon_c}{4 V_c}\left( 1 + 2k_c z \varphi  \right) \hat{a}^\dag \hat{a}. 
\end{equation}
From this term, we note that the light-interaction yields a constant reduction of the cavity resonant frequency $\omega_c$ of the form
\begin{equation}
\omega_c \rightarrow \tilde{\omega}_c = \omega_c \left( 1 - \frac{\epsilon_c V}{4 V_c}  \right). 
\end{equation}
The first-order correction in $z$ can now be quantised by promoting $z$ to an operator $z \rightarrow \hat{z} = \sqrt{\frac{\hbar}{2\omega_m m}} (\hat{b}^\dag + \hat{b})$ so that we find the interaction term
\begin{equation}
\hat{H}_{int} = -\omega_c \frac{V \epsilon_c}{2 V_c}  k_c \varphi \hat{z}. 
\end{equation}
We now use the fact that $k_c = \frac{\omega_c}{c}$ to write
\begin{equation}
\hat{H}_{int} = - \sqrt{\frac{\hbar}{2 \omega_m m }}\frac{ \omega_c^2 V \epsilon_c \varphi }{2 V_c c} \, \hat{a} ^\dag \hat{a} \left( \hat{b}^\dag + \hat{b} \right), 
\end{equation}
where we can define the final expression for the light--matter coupling:
\begin{equation}
g = \sqrt{\frac{\hbar}{2 \omega_m m }} \frac{ \omega_c^2 V \epsilon_c \varphi }{2 V_cc}. 
\end{equation}
In all traps, optical and Paul traps, the bead is trapped in a minimum of the potential. This occurs at $\varphi = \frac{\pi}{2}$. 

In optical traps, we can now modulate $\varphi \rightarrow \varphi(t)$, to change the light--matter coupling. If we let $\varphi(t) = \frac{\pi}{2} \left( 1 + \epsilon \sin{\omega_0 \, t} \right)$, we obtain the scenario we investigate in Section~\ref{sec:time:dependent:coupling}. Finally, we note that there might be many additional ways in which the coupling can be modulated that we have not discussed in this work.

\end{document}